# Title page

## Title

Fraction-variant VMAT planning for patients with complex gynecological and head-and-neck cancer

## Authors

Nathan Torelli [a]

[a] Department of Radiation Oncology, University Hospital Zurich and University of Zurich, Zurich, Switzerland

Madalyne Day [a]

[a] Department of Radiation Oncology, University Hospital Zurich and University of Zurich, Zurich, Switzerland

Jan Unkelbach [a]

[a] Department of Radiation Oncology, University Hospital Zurich and University of Zurich, Zurich, Switzerland

## Corresponding author

Nathan Torelli [a]

[a] Department of Radiation Oncology, University Hospital Zurich and University of Zurich, Zurich, Switzerland

Mail: Rämistrasse 100, 8091 Zurich, Switzerland

Email: nathan.torelli@usz.ch

Phone: +41 79 569 15 34

# Abstract

## Background and Purpose

Increasing the number of arcs in volumetric modulated arc therapy (VMAT) allows for better intensity modulation and may improve plan quality. However, this leads to longer delivery times, which may cause patient discomfort and increase intra-fractional motion. In this study, it was investigated whether the delivery of different VMAT plans in different fractions may improve the dosimetric quality and delivery efficiency for the treatment of patients with complex tumor geometries.

## Materials and Methods

A direct aperture optimization algorithm was developed which allows for the simultaneous optimization of different VMAT plans to be delivered in different fractions, based on their cumulative physical dose. Each VMAT plan is constrained to deliver a uniform dose within the target volume, such that the entire treatment does not alter the fractionation scheme and is robust against inter-fractional setup errors. This approach was evaluated in-silico for ten patients with gynecological and head-and-neck cancer.

## Results

For all patients, fraction-variant treatments achieved better target coverage and reduced the dose to critical organs-at-risk compared to fraction-invariant treatments that deliver the same plan in every fraction, where the dosimetric benefit was shown to increase with the number of different plans to be delivered. In addition, 1-arc and 2-arc fraction-variant treatments could approximate the dosimetric quality of 3-arc fraction-invariant treatments, while reducing the delivery time from 180 s to 60 s and 120 s, respectively.

## Conclusions

Fraction-variant VMAT treatments may achieve excellent dosimetric quality for patients with complex tumor geometries, while keeping the delivery time per fraction viable.

# 1. Introduction

In current clinical practice, most radiotherapy treatments are delivered over the course of several fractions. Although adaptive radiotherapy is being increasingly adopted to account for possible inter-fractional anatomical variations [1-3], the majority of the treatments nowadays still deliver the same dose distribution in every fraction, which has been optimized based on the planning CT. The dosimetric quality of such treatments highly depends on the delivery parameters which are selected for the treatment plan optimization, such as the number of different beam orientations [4] and number of apertures or arcs [5-6]. For instance, intensity modulated radiotherapy (IMRT) performs better when a larger number of beam orientations and a larger number of multileaf-collimated apertures are allowed [4,7], as a better intensity modulation can be achieved. Analogously, dual-arc volumetric modulated arc therapy (VMAT) reported better dosimetric results compared to single-arc VMAT treatments [5-6].

However, increasing the number of apertures or arcs in a treatment requires a longer delivery time. This is particularly critical for patients with large or complex tumor geometries, such as gynecological cancer with nodal involvement or bilateral head-and-neck cancer. For these patients, three or more arcs may be needed to achieve acceptable dosimetric quality, which may cause patient discomfort and increase the risk for intra-fractional motion. Technological advances such as flattening filter free (FFF) delivery [8] or high-speed multileaf collimators (MLC) [9] could only partly compensate for such longer delivery times. In that regard, the development of new methods to better balance the dosimetric quality of a treatments and its delivery efficiency are warranted.

Few approaches to address to improve the trade-off between plan quality and delivery efficiency have already been proposed in previous studies. O'Connor *et al* [10] and Gu *et al* [11], for example, suggested to deliver different beam orientations in different fractions. They showed for both non-coplanar intensity modulated radiotherapy (IMRT) and intensity modulated proton therapy (IMPT) that in this manner it is possible to use overall more beam orientations over the course of a treatment, but without increasing the delivery time per fraction. Such so-called fraction-variant treatments were shown to either improve the dosimetric quality of a treatment while keeping the delivery time per fraction viable, or alternatively to maintain a good plan quality while lowering the delivery time. Similar approaches were also investigated by Rossi *et al* [12] and by Torelli and Unkelbach [13] for stereotactic radiotherapy of prostate, liver metastases and large arteriovenous malformations.

In this work, we expand on previous studies by proposing an approach to improve the intensity modulation without increasing the delivery time. The approach consists of delivering different VMAT plans (defined through different aperture shapes and MU weights) in different fractions. In this way, overall more apertures can be used to approximate the optimal fluence, but without increasing the number of arcs per fraction and thus keeping the delivery time viable. Alternatively, a similar dosimetric plan quality may be achieved as a conventional fraction-invariant treatment while using fewer arcs per fraction. This approach, which is referred to as *fraction-variant VMAT planning*, is demonstrated in this work for five patients with large gynecological tumors and five patients with bilateral head-and-neck cancer.

## 2. Materials and methods

### 2.1. Patients

Fraction-variant VMAT planning has been retrospectively evaluated for five patients with gynecological cancer and five patients with head-and-neck cancer (Table 1). The patients with gynecological cancer have complex tumor geometries which also include the inguinal and pelvic lymph nodes, while the patients with head-and-neck cancer present very large bilateral tumors. For these reasons, for all selected patients 3 coplanar arcs were used in the original clinical VMAT treatment to deliver each fraction.

*Table 1: Characteristics of the 5 patients with gynecological cancer and the 5 patients with head-and-neck cancer considered in this study. For the patients with head-and-neck cancer, the target volumes and prescribed doses for PTV1, PTV2 and PTV3 are reported, where PTV3 ⊂ PTV2 ⊂ PTV1.*

|  |  | Tumor site | PTV volume (cc) | Prescribed dose (Gy) | Number of fractions |
|---|---|---|---|---|---|
| **Gynecological cancer** | **Patient 1** | Vulva | 2299 | 52.8 | 24 |
|  | **Patient 2** | Cervix | 1061 | 50.4 | 28 |
|  | **Patient 3** | Vulva | 2056 | 45[1] | 25 |
|  | **Patient 4** | Cervix | 1266 | 45 | 25 |
|  | **Patient 5** | Ovary | 1029 | 35 | 14 |
| **Head-and-neck cancer** | **Patient 6** | Tongue | 739 / 352 / 232 | 54 / 60 / 70 | 35 |
|  | **Patient 7** | Mouth | 629 / 453 / 248 | 54 / 60 / 70 | 35 |
|  | **Patient 8** | Mouth | 255 / 125 | 54 / 68 | 35 |
|  | **Patient 9** | Tonsil | 865 / 475 / 270 | 54 / 60 / 70 | 35 |
|  | **Patient 10** | Nasopharynx | 741 / 392 / 92 | 54 / 60 / 70 | 35 |

---

[1] A total dose of 57.5 Gy was prescribed to a boost region for patient 3.

## 2.2. Optimization problem for fraction-variant VMAT planning

Fraction-variant VMAT treatments are obtained by simultaneously optimizing $n \in \mathbb{N}$ different VMAT plans each to be delivered in $N/n \in \mathbb{N}$ distinct fractions (where $N \in \mathbb{N}$ is the total number of fractions), based on their cumulative physical dose. Formally, the optimization problem for fraction-variant VMAT planning reads:

$$\underset{\omega, (L, R)}{\text{minimize}} \quad f(\boldsymbol{d}) \tag{1}$$

$$\text{subject to} \quad d_i = \sum_{t=1}^{n} d_{it} \quad \forall i \tag{2}$$

$$d_{it} = \sum_{k \in K_t} \omega^k A_{it}^k (L_t^k, R_t^k) \quad \forall i, \forall t \tag{3}$$

$$\omega^k \geq 0 \quad \forall k \tag{4}$$

where $f$ is an objective function evaluated for the cumulative dose distribution $\boldsymbol{d}$, $d_i$ is the cumulative dose to voxel $i$, $d_{it}$ is the dose delivered to voxel $i$ in plan $t$ by all apertures $k \in K_t$ belonging to that plan, $A_{it}^k$ is the dose delivered to voxel $i$ by aperture $k$ in plan $t$ (which depends on the set of left and right MLC leaf positions $(L_t^k, R_t^k)$), and $\omega_k$ defines the MU weight of aperture $k$. Each aperture $k \in K_t$ is subject to constraints on the leaf positions as described in [14].

To guarantee that the treatment does not alter the fractionation scheme substantially and is robust against inter-fractional setup errors, each VMAT plan is required to deliver a similar prescribed dose to the target volume in each fraction. To this end, the objective function in Eq. (1) is defined as follows:

$$f(\boldsymbol{d}) = \sum_{i \in PTV} \omega_u (d_{min} - d_i)_+^2 + \omega_o (d_i - d_{max})_+^2 \tag{5}$$

$$+ \sum_{t=1}^{n} \sum_{i \in PTV} \omega_u \left(\frac{d_{min}}{n} - d_{it}\right)_+^2 + \omega_o \left(d_{it} - \frac{d_{max}}{n}\right)_+^2 \tag{6}$$

$$+ f_{OAR}(\boldsymbol{d}) \tag{7}$$

$$+ f_{NT}(\boldsymbol{d}) \tag{8}$$

Here, $PTV$ refers to the set of voxels belonging to the planning target volume, $OAR$ refers to the set of voxels belonging to the organs-at-risk and and $NT$ refers to the set of voxels belonging to normal tissue. The parameters $\omega_u$ and $\omega_o$ define the weights for the quadratic penalty functions used to penalize under- and over-dosage in the PTV, whereas $d_{min}$ and $d_{max}$ are the minimum and maximum prescribed doses to the PTV, respectively. The objective function $f$ consists of the following different terms. The planning objective in Eq. (5) requires that the prescribed tumor dose is homogeneously delivered to the PTV over the entire treatment, whereas the planning objective in Eq. (6) requires that the same fractional dose contribution is homogeneously delivered to the PTV in each plan $t$. The planning objectives in Eq. (7)-(8) are patient-specific terms to control the dose in the OARs and normal tissue, respectively (as further detailed below).

### 2.3. Optimization algorithm

The optimization problem described in Equations (1)-(4) is solved using a direct aperture optimization (DAO) algorithm, which has been developed into our in-house research treatment planning system and allows for the simultaneous optimization of multiple VMAT plans to be delivered in different fractions. The workflow of the DAO algorithm is illustrated in Figure 1. It combines a column generation approach to iteratively add promising multileaf collimated apertures along the VMAT arcs in the different plans and a gradient based DAO approach that optimizes the weights and refines the shapes of all the already added apertures at each iteration. Further details on the DAO algorithm are reported in the work of Torelli and Unkelbach [13] and in the Supplementary material (Appendix A).

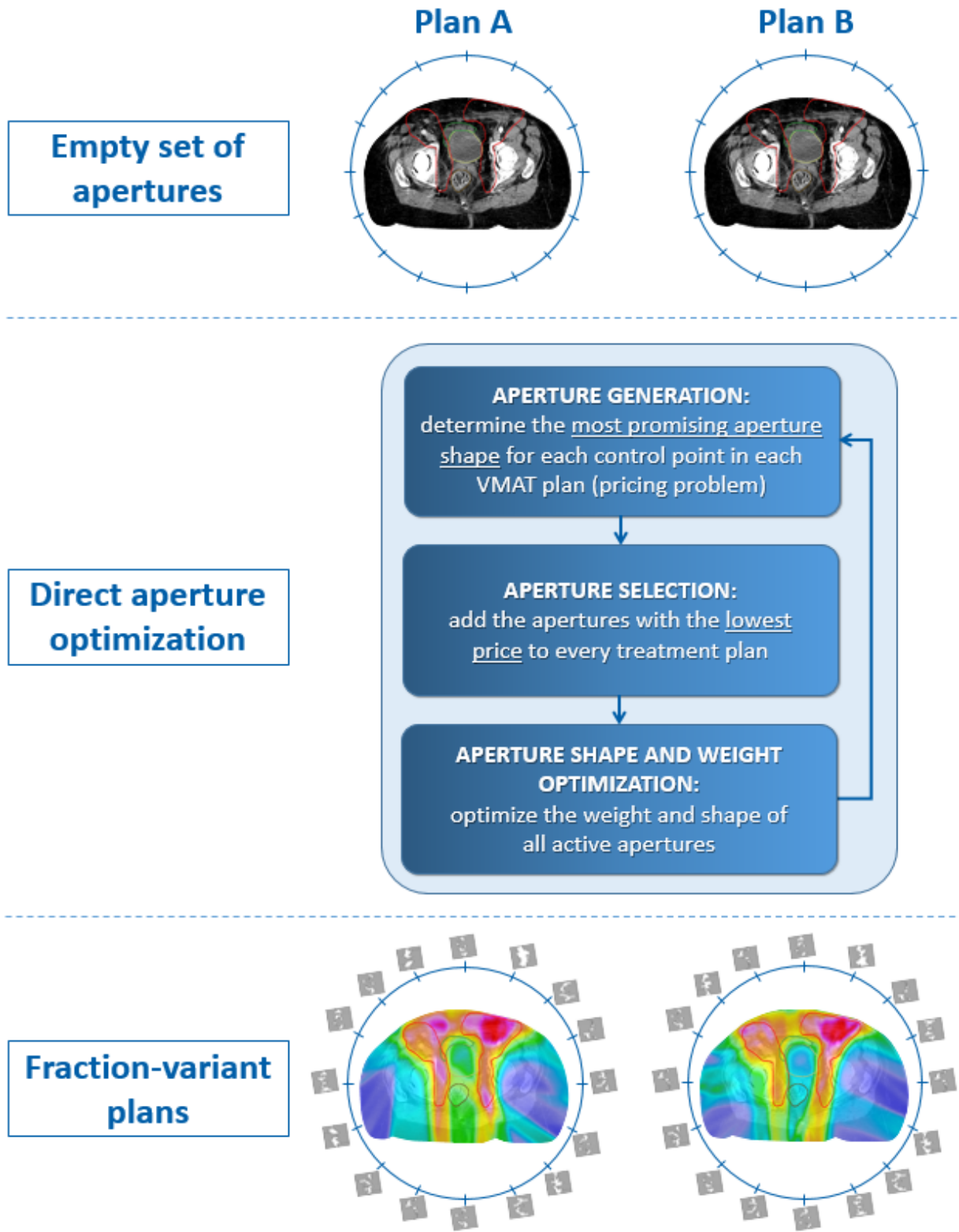

*Figure 1: Schematic illustration of the direct aperture optimization algorithm used to generate fraction-variant VMAT treatments. Starting from an empty set of control points for all plans, promising apertures at different control points are iteratively added to each plan (using a column generation approach), until apertures for all control points are defined. At each iteration, a gradient-based approach is used to adjust the weight and shapes of all already added apertures.*

## 2.4. Planning of fraction-variant VMAT treatments for patients with gynecological and head-and-neck cancer

For each patient, multiple 1-arc and 2-arc *fraction-variant* VMAT treatments were generated in-silico, each delivering a different number $n \in \left\{1, \ldots, N \middle| \frac{N}{n} \in \mathbb{N}\right\}$ of distinct VMAT plans. These treatments have then been benchmarked against *fraction-invariant* 1-arc, 2-arc and 3-arc VMAT treatments, respectively, that deliver the same dose distribution in every fraction, in terms of both dosimetric quality and delivery efficiency. All treatments have been optimized for a weighted sum of dose-volume, mean dose and normal tissue objectives. Similar objective parameters and priorities have been used as for the corresponding clinical plans, which have been generated in the commercial treatment planning system Eclipse (Varian, A Siemens Healtineers Company) by expert dosimetrists at our institution. The specific planning objectives and priorities used for each patient, as well as other details regarding the optimization parameters used in this study, are reported in the Supplementary material (Appendix B). For the fraction-variant treatments, the additional planning objectives defined in Equation (6) have been used to require that a similar dose is delivered to the PTV in every fraction.

Given that different dose distributions are delivered in different fractions, a robustness analysis has also been performed to investigate the sensitivity of fraction-variant VMAT treatments against inter-fractional setup errors.

## 3. Results

### 3.1. Dosimetric evaluation of fraction-variant VMAT treatments

Fraction-variant treatments delivered quite different plans in different fractions (Figures 2a-d). Each individual plan was sub-optimal in terms of target coverage and dose conformity, with high doses delivered to different parts of the normal tissue. However, the cumulative treatment resulting from the sum of all the individual plan achieved a much better dosimetric quality (Figure 2e). In particular, the cumulative treatment presented a more homogeneous target dose and maintained excellent OAR sparing (Figure 2f).

For patient 1, for example, a fraction-variant treatment that delivers 4 different plans in 7 fractions each was shown to reduce the mean bladder dose from 15.9 Gy to 13.7 Gy (-12.2%), the mean bowel dose from 15.6 Gy to 14.5 Gy (-7.1%) and the mean rectum dose from 20.9 Gy to 20.1 Gy (-3.8%) compared to a treatment that delivers the same 2-arc VMAT plan in every fraction (Figure 2g). At the same time, the homogeneity index ($HI = D_{90\%}/D_{10\%}$) within the PTV increased from 0.73 to 0.78 (+6.8%) and the conformity index ($CI = (V_{PTV} \cap V_{95\%})^2/(V_{PTV} V_{95\%})$) increased from 0.66 to 0.69. Also, the fraction-variant treatment could approximate the original 3-arc fraction-invariant VMAT treatment (Figure 2h), while reducing the delivery time per fraction from 180 s to 120 s. The original 3-arc fraction-invariant treatment, in fact, achieved similar or slightly better mean doses to the bladder, bowel and rectum of 12.8 Gy, 13.4 Gy and 17.1 Gy, respectively, but at the cost of poorer homogeneity index of 0.74 and conformity index of 0.68.

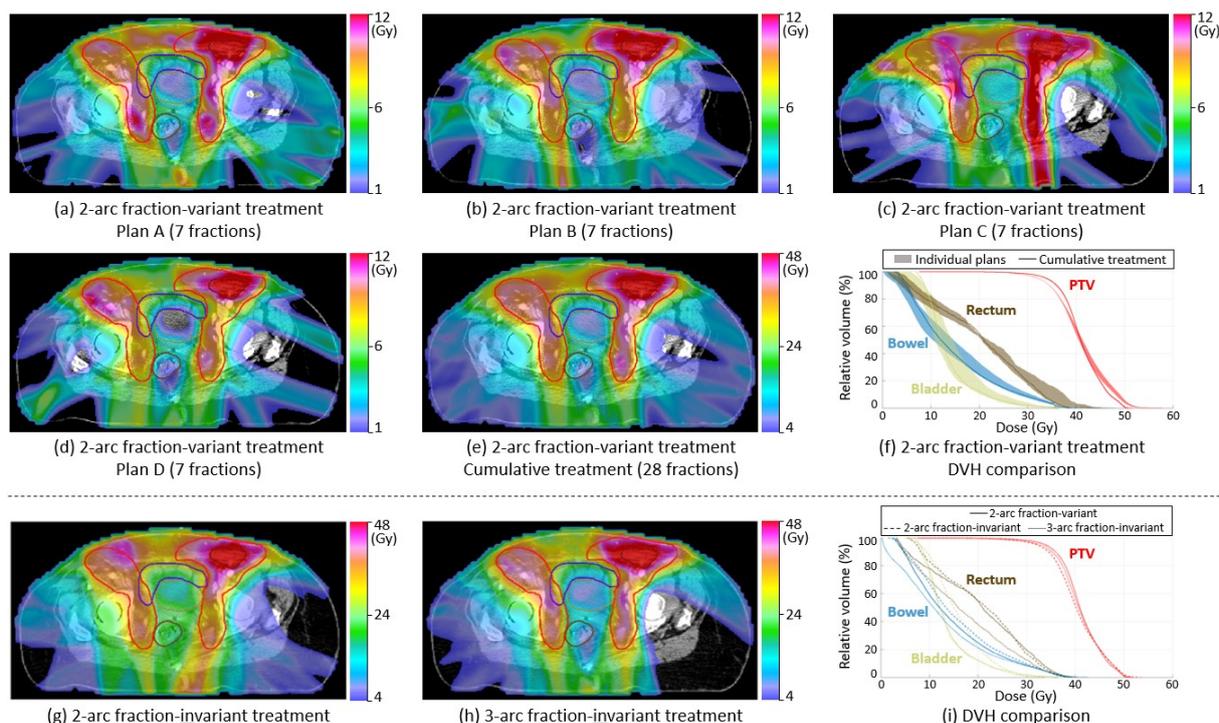

*Figure 2: Dose distributions and dose-volume histograms (DVH) obtained for patient 1 with (a-d) a fraction-variant treatment that delivers 4 different 2-arc VMAT plans in 6 fractions each, and with (g) 2-arc and (h) 3-arc fraction-invariant treatments. The dose-volume histogram (f) compares the cumulative treatment resulting from the sum of all individual plans (solid line) with the worst and best case scenarios of the individual plans (assuming that each plan was delivered over 24 fractions; shown by the DVH band). The dose-volume histogram (i) compares the fraction-variant and fraction-invariant treatments.*

Similar results were obtained also for patients 2-5 (Figure 3) and patients 6-10 (Figure 4), whose dose distributions and dose-volume histograms are reported in the Supplementary material (Appendix C). Overall, fraction-variant treatments achieved better target coverage and reduced the dose to critical OARs compared to the corresponding fraction-invariant treatment. Greater improvements were observed when the number of distinct plans was increased. For instance, a fraction-variant treatment that delivered a different 2-arc VMAT plan in every fraction reduced the mean doses to bladder, bowel and rectum in patient 1 down to 6.0 Gy, 10.1 Gy and 16.3 Gy, respectively, while maintaining an excellent homogeneity index of 0.78. For all patients, 1-arc and 2-arc fraction-variant treatments could approximate or even outperform the dosimetric quality of the original 3-arc fraction-invariant treatment, while significantly reducing the delivery time per fraction from 180 s to 60 s and 120 s, respectively.

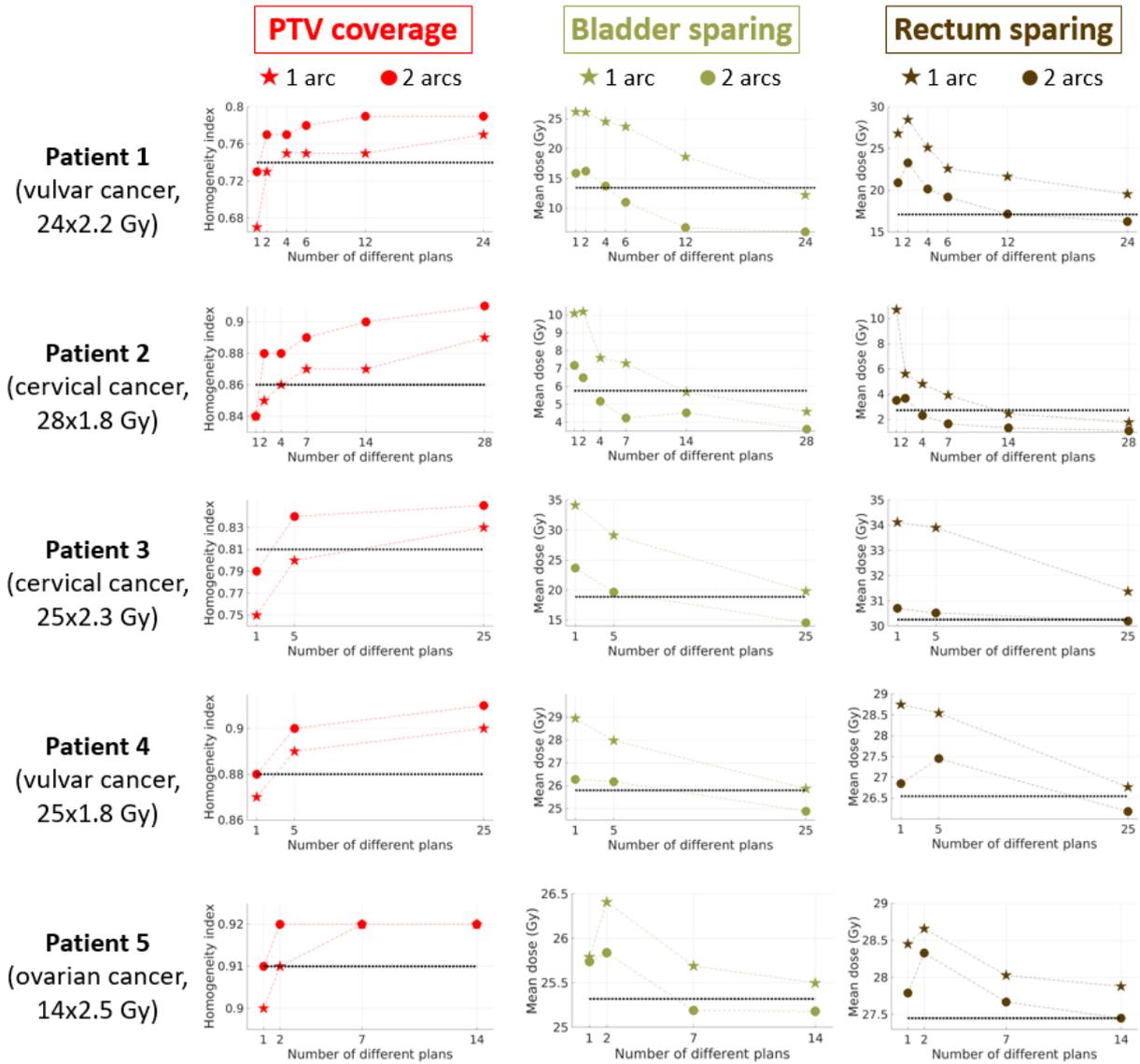

*Figure 3: Dosimetric results obtained for the patients with gynecological cancer with 1-arc fraction-variant VMAT treatments (denoted by a star marker) and 2-arc fraction-variant VMAT treatments (denoted by a round marker), as a function of the number $n_p$ of different plans generated. Each plan is meant to be delivered over $N/n_p$ fractions, where $N$ corresponds to the total number of fractions). The results for a 3-arc fraction-invariant VMAT treatment are reported by the black dashed line. Homogeneity index in the PTV is expressed as $HI = D_{90\%}/D_{10\%}$.*

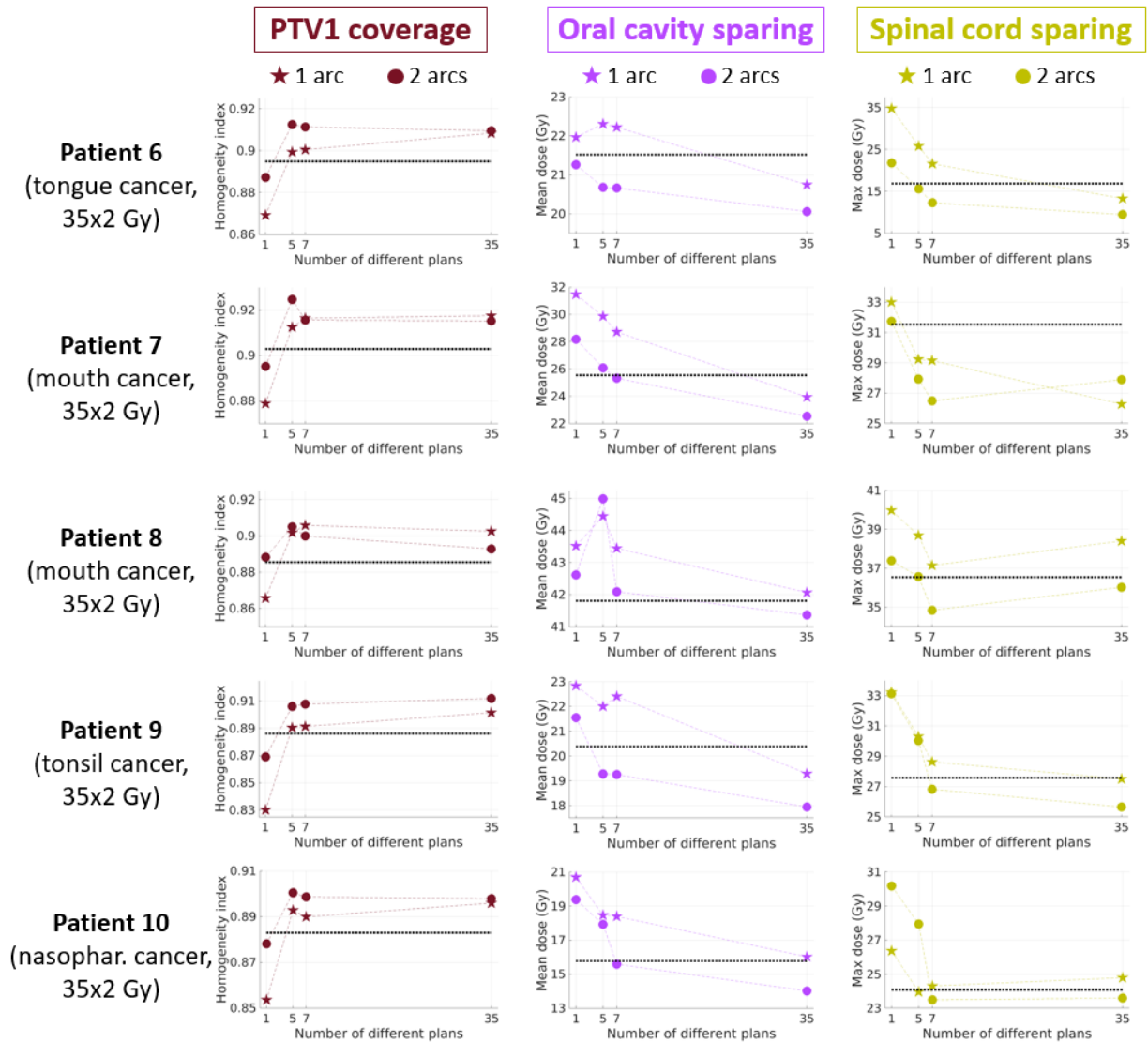

*Figure 4: Dosimetric results obtained for the patients with head-and-neck cancer with 1-arc fraction-variant VMAT treatments (denoted by a star marker) and 2-arc fraction-variant VMAT treatments (denoted by a round marker), as a function of the number $n_p$ of different plans generated. Each plan is meant to be delivered over $N/n_p$ fractions, where $N$ corresponds to the total number of fractions). The results for a 3-arc fraction-invariant VMAT treatment are reported by the black dashed line. Homogeneity index in the PTV is expressed as $HI = D_{90\%}/D_{10\%}$.*

## 3.2. Robustness of fraction-variant VMAT treatments against inter-fractional setup errors

The dosimetric results reported above assumed that every fractional dose distribution was delivered as planned. To determine the impact of random setup errors on the CTV dose coverage when different dose distributions are delivered in distinct fractions, a robustness analysis was performed (Figure 5). Because each VMAT plan in a fraction-variant treatment was enforced to deliver a similar prescribed dose to all parts of the PTV, fraction-variant VMAT treatments maintained a good CTV dose coverage also in the presence of setup errors. The sensitivity to setup errors was similar as for the corresponding fraction-invariant treatments.

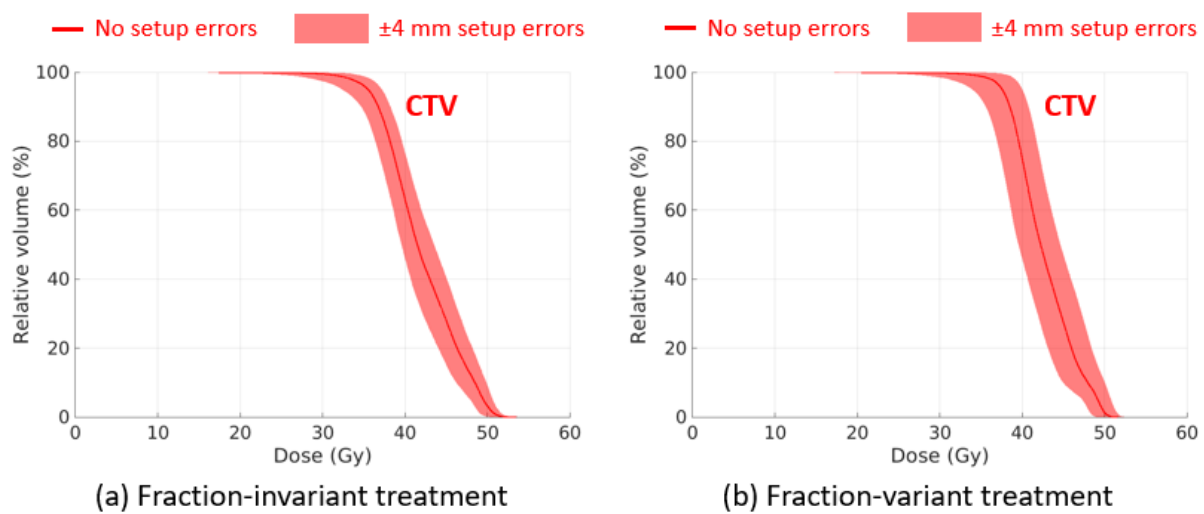

*Figure 3: Dose-volume histograms for the CTV in patient 1 evaluated for the scenario with no setup errors (solid line) and for multiple errors scenarios that assume ±4 mm setup errors in left–right, superior–inferior, and anterior–posterior directions in between the different fractions (bands). The magnitude of the setup errors is selected to approximately agree with the CTV-to-PTV margin expansion.*

## 4. Discussion

In current clinical practice, the same plan is delivered in all fractions. This limits the number of apertures/arcs that can be used, and consequently also the achievable dosimetric quality. In this study, it was demonstrated that by delivering different plans in different fractions, which are simultaneously optimized based on their cumulative dose distributions, it is possible to achieve an improved target coverage and better spare the OARs, without increasing the delivery time per fraction. This is due to the fact that overall more apertures can be used to approximate the ideal fluence. Alternatively, the proposed method can also be used to maintain a similar plan quality as with a conventional fraction-invariant treatment while reducing the delivery time per fraction. This may have a positive impact on patient comfort and limit intra-fraction motion uncertainties, in particular for patients with large or complex tumor geometries who may otherwise require three or more VMAT arcs to achieve an acceptable plan quality.

The delivery of different plans in distinct fractions has already been proposed by some authors [15-17]. López Alfonso *et al*, for example, proposed a concept named temporally feathered radiotherapy [15], where different OARs receive a slightly higher dose in few fractions and are better spared in the other fractions. The authors claimed that this may lead to lower OAR toxicities. Unkelbach *et al* suggested to treat different regions of the target volume in distinct fractions, while delivering a similar dose bath to the normal tissue in all fractions [16-17]. Such a concept, which is referred to as spatiotemporal fractionation, optimally exploits the fractionation effect by achieving partial tumor hypofractionation along with more uniform fractionation of the dose in the normal tissue. However, in most of those studies the main rationale for delivering different dose distributions in different fractions relied on radiobiological principles and was done deliberately. The fraction-variant VMAT planning approach proposed in this study, instead, complements previous research on fraction-variant radiotherapy [10-12], in which the dosimetric benefit is solely motivated by the use of additional degrees of freedom in between different fractions.

Though fraction-variant VMAT planning has been shown to be a promising approach for improving dosimetric quality and/or delivery efficiency, the following aspects must be carefully considered. First, no commercially available treatment planning system currently supports the simultaneous optimization of multiple dose distributions. A possible solution to implement fraction-variant VMAT planning in the clinic would be to use sequential optimization, where a single VMAT plan is first generated for a patient and subsequently additional VMAT plans are optimized on top of each other to be delivered in only a part of the fractions. Second, both the planning time and the time needed for quality assurance increase linearly with the number of different plans to be generated. However, this is unlikely to be an insurmountable difficulty. The continuous development in computational hardware, in fact, makes optimization increasingly faster, while the need for plan-specific quality assurance may be reduced through the use of artificial intelligence [18].

Future work may be performed to also integrate radiobiological parameters in the treatment plan optimization problem. Also, fraction-variant VMAT planning may be combined with fraction-variant beam orientation optimization or collimator angle optimization to exploit even more degrees of freedom over the treatment course.

# Appendix A    Optimization algorithm

In this section, further details are provided of the DAO algorithm that has been used in this study to solve the fraction-variant VMAT planning problem, whose workflow is illustrated in Figure 1.

The DAO algorithm starts by assuming an empty set of apertures for each of the $n$ different VMAT plans in a fraction-variant treatment. Promising multileaf collimated apertures are then generated for each control point in each plan by solving a so-called pricing problem [1], which minimizes the first-order perturbation for each bixel on the objective function. The resulting candidate apertures have an associated price, given by the sum of the gradient contributions of each of the bixels not covered by the MLC, and the apertures with the lowest price in each distinct plan are added to the treatment plan at each iteration. During the first $\lambda \in \{1, \dots, n\}$ iterations of the column generation method, only one single aperture is added to the VMAT plan $t = \lambda$. This is necessary to allow for the dose distributions delivered in the different plans to possibly diverge, as the gradient information for all plans with an empty pool of apertures is the same. This process is iteratively repeated until all control points along the VMAT arcs of each individual plan are populated with an aperture.

After every iteration of the column generation approach, both the MU weights and the shapes of all the already defined apertures are refined using a gradient based DAO approach, aiming to minimize an objective function $f(\boldsymbol{d})$ evaluated for the cumulative physical dose $\boldsymbol{d}$, as described in Eq. (1). The proposed gradient based DAO approach, which is inspired by the work of Cassioli and Unkelbach [2], exploits gradient information to locally optimize the positions of the MLC leaves and the MU weights in an iterative way. At each iteration of the gradient based DAO method, a restricted optimization problem is solved, where the left and right MLC leaf positions $x_{L,l}^{kt} \in L_t^k$ and $x_{R,l}^{kt} \in R_t^k$ for all leaf pairs $l \in \{1, \dots, |L_t^k|\}$ and for all apertures $k \in K_t$ are constrained to the bixels $b_{L,l}^{kt}$ and $b_{R,l}^{kt}$ they are positioned in. Under this assumption, the dose $A_{it}^k$ delivered to voxel $i$ by aperture $k$ in plan $t$ (Eq. (3)) can be decomposed into the following separate terms:

$$A_{it}^k = \sum_{l=1}^{|L_t^k|} \left[ \delta_{L,l}^{kt} D_{ib_{L,l}^{kt}} + \delta_{R,l}^{kt} D_{ib_{R,l}^{kt}} + \sum_{j=b_{L,l}^{kt}+1}^{b_{R,l}^{kt}-1} D_{ij} \right] \qquad (A.1)$$

Here, $\delta_{L,l}^{kt} \in [0,1]$ and $\delta_{R,l}^{kt} \in [0,1]$ describe the fractional opening of the bixels that the left and right MLC leaves in leaf pair $l$ are restricted to, and $D_{ij}$ is the dose-influence matrix term storing the dose contribution of bixel $j$ to voxel $i$ per unit intensity. As the dose contribution is linear to the fraction of the bixel area open, this leads to a well-behaved optimization problem for the leaf positions and the aperture weights, which can be solved efficiently using standard gradient-based algorithms. In this study, our in-house implementation of the L-BFGS quasi-Newton method [3] was used, where the constraints $\delta_L^{kl} \in [0,1]$ and $\delta_R^{kl} \in [0,1]$ are handled by projection methods.

If a MLC leaf is moved at the edge of a bixel after an iteration of the gradient based DAO method, its position can be confined to the neighboring bixel in the next iteration (details on the leaf-to-bixel assignment procedure are discussed in the work of Cassioli and Unkelbach [2]). This allows for larger MLC leaf displacements over multiple iterations. In this work, the iterative refinement of the MLC leaf positions and the MU weights stops either after 20 iterations of the gradient based DAO algorithm or sooner if the decrease in the relative value of the objective function between two consecutive iterations is smaller than $\epsilon = 10^{-6}$.

## References Appendix A

[1] *Romeijn, H. Edwin et al. "A Column Generation Approach to Radiation Therapy Treatment Planning Using Aperture Modulation." SIAM J. Optim. 15 (2005): 838-862.*

[2] *Cassioli A, Unkelbach J. Aperture shape optimization for IMRT treatment planning. Phys Med Biol. 2013;58(2):301-318. doi:10.1088/0031-9155/58/2/301.*

[3] *Wright SJ, Nocedal J. Numerical Optimization. Vol 2. Springer;1999.*

# Appendix B  Patient-specific optimization parameters

In this section, the planning objectives and optimization parameters used to generate the fraction-invariant and fraction-variant treatments are detailed for all patients.

## B.1  Patient 1

For patient 1, the objective function in Equation (1) reads as follows:

$$f(\mathbf{d}) = \frac{1}{|PTV1|} \sum_{i \in PTV1} 150\,(52.8 - d_i)_+^2 + 160\,(d_i - 53.3)_+^2 \tag{B1.1}$$

$$+ \frac{1}{|PTV2|} \sum_{i \in PTV2} 150\,(42.4 - d_i)_+^2 + 100\,(d_i - 50)_+^2 \tag{B1.2}$$

$$+ \frac{1}{|AC|} \sum_{i \in AC} 75\,(d_i - 38)_+^2 + 8\,d_i \tag{B1.3}$$

$$+ \frac{1}{|BD|} \sum_{i \in BD} 70\,(d_i - 38)_+^2 + 75\,d_i \tag{B1.4}$$

$$+ \frac{1}{|BW|} \sum_{i \in BW} 75\,(d_i - 38)_+^2 + 85\,d_i \tag{B1.5}$$

$$+ \frac{1}{|FHL|} \sum_{i \in FHL} 80\,d_i \tag{B1.6}$$

$$+ \frac{1}{|RC|} \sum_{i \in RC} 75\,(d_i - 32)_+^2 + 80\,d_i \tag{B1.7}$$

$$+ \frac{1}{|SC|} \sum_{i \in SC} 75\,(d_i - 42)_+^2 + 80\,d_i \tag{B1.8}$$

$$+ \frac{1}{|NT|} \sum_{i \in NT} 140\,(d_i - d_i^{max})_+^2 \tag{B1.9}$$

where $PTV1$ denotes the set of voxels belonging to the PTV1, $PTV2$ denotes the set of voxels belonging to the PTV2, and $AC, BD, BW, FHL, RC, SC$ and $NT$ are the set of voxels belonging to the anal canal, bladder, bowel, left femur head, rectum, sigmoid colon and normal tissue, respectively. The planning objective in Equation (B1.9) corresponds to the normal tissue objective (NTO) implemented in the Eclipse Treatment Planning System (Varian Medical Systems, Inc.), where $b_i^{max}$ is a voxel-dependent value defined as:

$$d_i^{max} = \begin{cases} d_0 & , if\ x_i < x_0 \\ d_0 e^{-\kappa(x_i - x_0)} + d_\infty\bigl(1 - e^{-\kappa(x_i - x_0)}\bigr) & , if\ x_i \geq x_0 \end{cases} \tag{B1.10}$$

Here, $x_i$ indicates the distance of a normal tissue voxel $i$ from the PTV edge. For the Equation (B1.10), the following parameters have been used: $d_0 = 50$ Gy, $d_\infty = 10$ Gy, $x_0 = 0.5$ cm and $\kappa = 0.6$ cm$^{-1}$. In addition to the planning objectives described in Equations (B1.1)-(B1.9), the planning objective in Equation (6) has been used to optimize the fraction-variant treatments for both the $PTV1$ and $PTV2$, using similar weights as for the objectives in Equations (B1.1) and (B1.2), respectively.

The bixel size is set to 5 x 5 mm² and the photon energy is 6 MV. A non-uniform dose grid size is used throughout the body, with small voxels of 4.7 x 4.7 x 2.0 mm³ in size that are used in the PTV and close to the PTV, where a larger dose gradient is expected. At a distance between 2 cm and 4 cm from the PTV edge, medium-size voxels are used with 8-fold volume, whereas at distances larger than 4 cm from the PTV edge large-size voxels are used with 64-fold volume. As previously shown by Mueller *et al* [1], the use of a non-uniform dose grid size allows to considerably enhance the computational efficiency with negligible trade-offs on the plan accuracy. All dosimetric results, however, are evaluated based on the finest (small) dose grid size.

## B.2   Patient 2

For patient 2, the objective function in Equation (1) reads as follows:

$$f(\boldsymbol{d}) = \frac{1}{|PTV1|} \sum_{i \in PTV1} 150\,(49.9 - d_i)_+^2 + 180\,(d_i - 51)_+^2 \qquad (B2.1)$$

$$+ \frac{1}{|AC|} \sum_{i \in AC} 140\,(d_i - 45)_+^2 + 100\,d_i \qquad (B2.2)$$

$$+ \frac{1}{|BD|} \sum_{i \in BD} 140\,(d_i - 20)_+^2 + 70\,d_i \qquad (B2.3)$$

$$+ \frac{1}{|BW|} \sum_{i \in BW} 140\,(d_i - 45)_+^2 + 80\,d_i \qquad (B2.4)$$

$$+ \frac{1}{|FHL|} \sum_{i \in FHL} 80\,(d_i - 30)_+^2 + 70\,d_i \qquad (B2.5)$$

$$+ \frac{1}{|FHR|} \sum_{i \in FHR} 80\,(d_i - 28)_+^2 + 70\,d_i \qquad (B2.6)$$

$$+ \frac{1}{|RC|} \sum_{i \in RC} 90\,(d_i - 20)_+^2 + 80\,d_i \qquad (B2.7)$$

$$+ \quad \frac{1}{|SC|} \sum_{i \in SC} 140 \, (d_i - 45)_+^2 + 80 \, d_i \tag{B2.8}$$

$$+ \quad \frac{1}{|NT|} \sum_{i \in NT} 150 \, (d_i - d_i^{max})_+^2 \tag{B2.9}$$

Here $FHR$ refers to the set of voxels belonging to the right femur head. For the NTO in Equation (B2.9), the following parameters have been used: $d_0 = 50$ Gy, $d_\infty = 10$ Gy, $x_0 = 0.5$ cm and $\kappa = 0.6$ cm$^{-1}$. In addition to the planning objectives described in Equations (B2.1)-(B2.9), the planning objective in Equation (6) has been used to optimize the fraction-variant treatments for the $PTV1$, using similar weights as for the objectives in Equations (B2.1).

Similar as for patient 1, a non-uniform dose grid size is used, where the size of the smaller voxels is 3.9 x 3.9 x 2.0 mm$^3$. The bixel size is set to 5 x 5 mm$^2$ and the photon energy is 6 MV.

### B.3   Patient 3

For patient 3, the objective function in Equation (1) reads as follows:

$$f(\boldsymbol{d}) \quad = \quad \frac{1}{|PTV1|} \sum_{i \in PTV1} 150 \, (56 - d_i)_+^2 + 180 \, (d_i - 60.4)_+^2 \tag{B3.1}$$

$$+ \quad \frac{1}{|PTV2|} \sum_{i \in PTV2} 150 \, (56 - d_i)_+^2 + 180 \, (d_i - 60.4)_+^2 \tag{B3.2}$$

$$+ \quad \frac{1}{|PTV3|} \sum_{i \in PTV3} 150 \, (44 - d_i)_+^2 + 150 \, (d_i - 50)_+^2 \tag{B3.3}$$

$$+ \quad \frac{1}{|AC|} \sum_{i \in AC} 70 \, (d_i - 45)_+^2 \tag{B3.4}$$

$$+ \quad \frac{1}{|BD|} \sum_{i \in BD} 70 \, d_i \tag{B3.5}$$

$$+ \quad \frac{1}{|FHL|} \sum_{i \in FHL} 70 \, (d_i - 40)_+^2 \tag{B3.6}$$

$$+ \quad \frac{1}{|FHR|} \sum_{i \in FHR} 70 \, (d_i - 40)_+^2 \tag{B3.7}$$

$$+ \quad \frac{1}{|RC|} \sum_{i \in RC} 70 \, (d_i - 45)_+^2 \tag{B3.8}$$

$$+ \quad \frac{1}{|SC|} \sum_{i \in SC} 110 \, (d_i - 40)_+^2 + 80 \, d_i \tag{B3.9}$$

$$+ \quad \frac{1}{|SB|} \sum_{i \in SB} 120\, (d_i - 41)_+^2 + 70\, d_i \qquad (B3.10)$$

$$+ \quad \frac{1}{|NT|} \sum_{i \in NT} 150\, (d_i - d_i^{max})_+^2 \qquad (B3.11)$$

Here $PTV3$ refers to the set of voxels belonging to the PTV3, while $SB$ is the set of voxels belonging to the small bowel. For the NTO in Equation (B3.11), the following parameters have been used: $d_0 = 41.8$ Gy, $d_\infty = 7.5$ Gy, $x_0 = 0.5$ cm and $\kappa = 0.6$ cm$^{-1}$. In addition to the planning objectives described in Equations (B3.1)-(B3.11), the planning objective in Equation (6) has been used to optimize the fraction-variant treatments for both the $PTV1$, $PTV2$ and $PTV3$, using similar weights as for the objectives in Equations (B3.1), (B3.2) and (B3.3), respectively.

A non-uniform dose grid size is used, where the size of the smaller voxels is 3.9 x 3.9 x 2.0 mm³. The bixel size is set to 5 x 5 mm² and the photon energy is 6 MV.

### B.4 Patient 4

For patient 4, the objective function in Equation (1) reads as follows:

$$f(\boldsymbol{d}) \quad = \quad \frac{1}{|PTV1|} \sum_{i \in PTV1} 180\, (43.5 - d_i)_+^2 + 300\, (d_i - 46.7)_+^2 \qquad (B4.1)$$

$$+ \quad \frac{1}{|BD|} \sum_{i \in BD} 88\, (d_i - 45)_+^2 + 88\, d_i \qquad (B4.2)$$

$$+ \quad \frac{1}{|BW|} \sum_{i \in BW} 88\, (d_i - 45)_+^2 + 88\, d_i \qquad (B4.3)$$

$$+ \quad \frac{1}{|FHL|} \sum_{i \in FHL} 77\, (d_i - 20)_+^2 \qquad (B4.4)$$

$$+ \quad \frac{1}{|FHR|} \sum_{i \in FHR} 77\, (d_i - 20)_+^2 \qquad (B4.5)$$

$$+ \quad \frac{1}{|RC|} \sum_{i \in RC} 50\, d_i \qquad (B4.6)$$

$$+ \quad \frac{1}{|SC|} \sum_{i \in SC} 50\, d_i \qquad (B4.7)$$

$$+ \quad \frac{1}{|BO|} \sum_{i \in BO} 66\, (d_i - 45)_+^2 + 66\, d_i \qquad (B4.8)$$

$$+ \quad \frac{1}{|NT|} \sum_{i \in NT} 150 \, (d_i - d_i^{max})_+^2 \qquad (B4.9)$$

Here $BO$ refers to the set of voxels belonging to the bones. For the NTO in Equation (B4.9), the following parameters have been used: $d_0 = 41.3$ Gy, $d_\infty = 7.5$ Gy, $x_0 = 0.5$ cm and $\kappa = 0.6$ cm$^{-1}$. In addition to the planning objectives described in Equations (B4.1)-(B4.9), the planning objective in Equation (6) has been used to optimize the fraction-variant treatments for the $PTV1$, using similar weights as for the objectives in Equations (B4.1).

A non-uniform dose grid size is used, where the size of the smaller voxels is 3.9 x 3.9 x 2.0 mm³. The bixel size is set to 5 x 5 mm² and the photon energy is 6 MV.

## B.5  Patient 5

For patient 5, the objective function in Equation (1) reads as follows:

$$f(\boldsymbol{d}) = \frac{1}{|PTV1|} \sum_{i \in PTV1} 200 \, (34 - d_i)_+^2 + 300 \, (d_i - 37)_+^2 \qquad (B5.1)$$

$$+ \quad \frac{1}{|BD|} \sum_{i \in BD} 88 \, (d_i - 31)_+^2 + 88 \, d_i \qquad (B5.2)$$

$$+ \quad \frac{1}{|FHL|} \sum_{i \in FHL} 88 \, (d_i - 17)_+^2 \qquad (B5.3)$$

$$+ \quad \frac{1}{|FHR|} \sum_{i \in FHR} 88 \, (d_i - 17)_+^2 \qquad (B5.4)$$

$$+ \quad \frac{1}{|RC|} \sum_{i \in RC} 110 \, (d_i - 33)_+^2 + 50 \, d_i \qquad (B5.5)$$

$$+ \quad \frac{1}{|NT|} \sum_{i \in NT} 150 \, (d_i - d_i^{max})_+^2 \qquad (B5.6)$$

For the NTO in Equation (B5.6), the following parameters have been used: $d_0 = 32.3$ Gy, $d_\infty = 5$ Gy, $x_0 = 0.5$ cm and $\kappa = 0.6$ cm$^{-1}$. In addition to the planning objectives described in Equations (B5.1)-(B5.6), the planning objective in Equation (6) has been used to optimize the fraction-variant treatments for the $PTV1$, using similar weights as for the objectives in Equations (B5.1).

A non-uniform dose grid size is used, where the size of the smaller voxels is 3.9 x 3.9 x 2.0 mm³. The bixel size is set to 5 x 5 mm² and the photon energy is 6 MV.

## B.6 Patient 6

For patient 6, the objective function in Equation (1) reads as follows:

$$f(\mathbf{d}) = \frac{1}{|PTV1|} \sum_{i \in PTV1} 130\,(53 - d_i)_+^2 + 125\,(d_i - 55.7)_+^2 \qquad (B6.1)$$

$$+ \frac{1}{|PTV2|} \sum_{i \in PTV2} 130\,(59 - d_i)_+^2 + 125\,(d_i - 63.2)_+^2 \qquad (B6.2)$$

$$+ \frac{1}{|PTV3|} \sum_{i \in PTV3} 135\,(69 - d_i)_+^2 + 124\,(d_i - 72)_+^2 \qquad (B6.3)$$

$$+ \frac{1}{|DT|} \sum_{i \in DT} 77\,(d_i - 43)_+^2 + 50\,d_i \qquad (B6.4)$$

$$+ \frac{1}{|B|} \sum_{i \in B} 88\,(d_i - 43)_+^2 + 50\,d_i \qquad (B6.5)$$

$$+ \frac{1}{|BS|} \sum_{i \in B} 116\,(d_i - 33)_+^2 + 66\,d_i \qquad (B6.6)$$

$$+ \frac{1}{|SC|} \sum_{i \in SC} 130\,(d_i - 36)_+^2 + 77\,d_i \qquad (B6.7)$$

$$+ \frac{1}{|OC|} \sum_{i \in OC} 77\,(d_i - 49)_+^2 + 88\,d_i \qquad (B6.8)$$

$$+ \frac{1}{|PL|} \sum_{i \in PL} 120\,d_i \qquad (B6.9)$$

$$+ \frac{1}{|PR|} \sum_{i \in PR} 100\,d_i \qquad (B6.10)$$

$$+ \frac{1}{|PX|} \sum_{i \in PX} 50\,(d_i - 50)_+^2 + 88\,d_i \qquad (B6.11)$$

$$+ \frac{1}{|NT|} \sum_{i \in NT} 120\,(d_i - d_i^{max})_+^2 \qquad (B6.12)$$

where $DT$, $B$, $BS$, $SC$, $OC$, $PL$, $PR$ and $PX$ denote the set of voxels belonging to the dorsal tissue, brain, brainstem, spinal cord, oral cavity, left parotid, right parotid and pharynx, respectively. For the NTO in Equation (B6.12), the following parameters have been used: $d_0 = 50$ Gy, $d_\infty = 10$ Gy, $x_0 = 0.5$ cm and $\kappa = 0.6$ cm$^{-1}$. In addition to the planning objectives described in Equations (B6.1)-(B6.12), the planning objective in Equation (6) has been used to optimize the fraction-variant treatments for $PTV1$, $PTV2$ and $PTV3$, using similar weights as for the objectives in Equations (B6.1), (B6.2) and (B6.3), respectively.

A non-uniform dose grid size is used, where the size of the smaller voxels is 3.9 x 3.9 x 2.0 mm³. The bixel size is set to 5 x 5 mm² and the photon energy is 6 MV.

## B.7 Patient 7

For patient 7, the objective function in Equation (1) reads as follows:

$$f(\boldsymbol{d}) = \frac{1}{|PTV1|} \sum_{i \in PTV1} 150 (53.5 - d_i)_+^2 + 150 (d_i - 57.4)_+^2 \qquad (B7.1)$$

$$+ \frac{1}{|PTV2|} \sum_{i \in PTV2} 160 (59.5 - d_i)_+^2 + 170 (d_i - 66)_+^2 \qquad (B7.2)$$

$$+ \frac{1}{|PTV3|} \sum_{i \in PTV3} 160 (69.5 - d_i)_+^2 + 160 (d_i - 73.5)_+^2 \qquad (B7.3)$$

$$+ \frac{1}{|B|} \sum_{i \in B} 100 (d_i - 47)_+^2 \qquad (B7.4)$$

$$+ \frac{1}{|BS|} \sum_{i \in B} 120 (d_i - 55)_+^2 + 110 \, d_i \qquad (B7.5)$$

$$+ \frac{1}{|SC|} \sum_{i \in SC} 120 (d_i - 35)_+^2 \qquad (B7.6)$$

$$+ \frac{1}{|OC|} \sum_{i \in OC} 120 (d_i - 59)_+^2 + 120 \, d_i \qquad (B7.7)$$

$$+ \frac{1}{|PL|} \sum_{i \in PL} 120 (d_i - 49)_+^2 + 120 \, d_i \qquad (B7.8)$$

$$+ \frac{1}{|PR|} \sum_{i \in PR} 120 (d_i - 42)_+^2 + 125 \, d_i \qquad (B7.9)$$

$$+ \frac{1}{|PX|} \sum_{i \in PX} 100 (d_i - 50)_+^2 + 50 \, d_i \qquad (B7.10)$$

$$+ \frac{1}{|NT|} \sum_{i \in NT} 120 (d_i - d_i^{max})_+^2 \qquad (B7.11)$$

For the NTO in Equation (B7.11), the following parameters have been used: $d_0 = 50$ Gy, $d_\infty = 10$ Gy, $x_0 = 0.5$ cm and $\kappa = 0.6$ cm$^{-1}$. In addition to the planning objectives described in Equations (B7.1)-(B7.11), the planning objective in Equation (6) has been used to optimize the fraction-variant treatments for $PTV1$, $PTV2$ and $PTV3$, using similar weights as for the objectives in Equations (B7.1), (B7.2) and (B7.3), respectively.

A non-uniform dose grid size is used, where the size of the smaller voxels is 3.9 x 3.9 x 2.0 mm³. The bixel size is set to 5 x 5 mm² and the photon energy is 6 MV.

## B.8 Patient 8

For patient 8, the objective function in Equation (1) reads as follows:

$$f(\mathbf{d}) = \frac{1}{|PTV1|} \sum_{i \in PTV1} 125\,(52.3 - d_i)_+^2 + 125\,(d_i - 55.6)_+^2 \qquad (B8.1)$$

$$+ \frac{1}{|PTV2|} \sum_{i \in PTV2} 125\,(66.6 - d_i)_+^2 + 140\,(d_i - 68.8)_+^2 \qquad (B8.2)$$

$$+ \frac{1}{|DT|} \sum_{i \in B} 50\,(d_i - 54)_+^2 + 50\,d_i \qquad (B8.3)$$

$$+ \frac{1}{|B|} \sum_{i \in B} 50\,(d_i - 33)_+^2 + 50\,d_i \qquad (B8.4)$$

$$+ \frac{1}{|SC|} \sum_{i \in SC} 125\,(d_i - 40)_+^2 + 77\,d_i \qquad (B8.5)$$

$$+ \frac{1}{|OC|} \sum_{i \in OC} 77\,d_i \qquad (B8.6)$$

$$+ \frac{1}{|PL|} \sum_{i \in PL} 120\,d_i \qquad (B8.7)$$

$$+ \frac{1}{|PR|} \sum_{i \in PR} 110\,d_i \qquad (B8.8)$$

$$+ \frac{1}{|PX|} \sum_{i \in PX} 88\,(d_i - 64)_+^2 + 99\,d_i \qquad (B8.9)$$

$$+ \frac{1}{|BPL|} \sum_{i \in BPL} 88\,(d_i - 63)_+^2 + 50\,d_i \qquad (B8.10)$$

$$+ \frac{1}{|BPR|} \sum_{i \in BPR} 50\,(d_i - 52)_+^2 + 50\,d_i \qquad (B8.11)$$

$$+ \frac{1}{|LX|} \sum_{i \in LX} 66\,(d_i - 63)_+^2 + 77\,d_i \qquad (B8.12)$$

$$+ \frac{1}{|MD|} \sum_{i \in MD} 77\,(d_i - 52)_+^2 + 88\,d_i \qquad (B8.13)$$

$$+ \frac{1}{|SML|} \sum_{i \in SML} 110\,d_i \qquad (B8.14)$$

$$+ \quad \frac{1}{|NT|} \sum_{i \in NT} 120 \, (d_i - d_i^{max})_+^2 \qquad (B8.15)$$

where $BPL$, $BPR$, $LX$, $MD$ and $SML$ denote the set of voxels belonging to the left brachial plexus, right brachial plexus, larynx, mandible and left submandibular gland, respectively. For the NTO in Equation (B8.15), the following parameters have been used: $d_0 = 50$ Gy, $d_\infty = 10$ Gy, $x_0 = 0.5$ cm and $\kappa = 0.6$ cm$^{-1}$. In addition to the planning objectives described in Equations (B8.1)-(B8.15), the planning objective in Equation (6) has been used to optimize the fraction-variant treatments for $PTV1$ and $PTV2$, using similar weights as for the objectives in Equations (B8.1) and (B8.2), respectively.

A non-uniform dose grid size is used, where the size of the smaller voxels is 3.9 x 3.9 x 2.0 mm$^3$. The bixel size is set to 5 x 5 mm$^2$ and the photon energy is 6 MV.

## B.9  Patient 9

For patient 9, the objective function in Equation (1) reads as follows:

$$f(\mathbf{d}) = \frac{1}{|PTV1|} \sum_{i \in PTV1} 130 \, (53 - d_i)_+^2 + 145 \, (d_i - 55.6)_+^2 \qquad (B9.1)$$

$$+ \quad \frac{1}{|PTV2|} \sum_{i \in PTV2} 120 \, (57.5 - d_i)_+^2 + 110 \, (d_i - 61.5)_+^2 \qquad (B9.2)$$

$$+ \quad \frac{1}{|PTV3|} \sum_{i \in PTV3} 135 \, (67.6 - d_i)_+^2 + 135 \, (d_i - 71.5)_+^2 \qquad (B9.3)$$

$$+ \quad \frac{1}{|DT|} \sum_{i \in B} 77 \, (d_i - 42)_+^2 + 50 \, d_i \qquad (B9.4)$$

$$+ \quad \frac{1}{|B|} \sum_{i \in B} 66 \, (d_i - 38)_+^2 + 66 \, d_i \qquad (B9.5)$$

$$+ \quad \frac{1}{|BS|} \sum_{i \in B} 88 \, (d_i - 32)_+^2 + 50 \, d_i \qquad (B9.6)$$

$$+ \quad \frac{1}{|SC|} \sum_{i \in SC} 119 \, (d_i - 38)_+^2 + 77 \, d_i \qquad (B9.7)$$

$$+ \quad \frac{1}{|OC|} \sum_{i \in OC} 88 \, (d_i - 54)_+^2 + 88 \, d_i \qquad (B9.8)$$

$$+ \quad \frac{1}{|PL|} \sum_{i \in PL} 99 \, d_i \qquad (B9.9)$$

$$+ \quad \frac{1}{|PR|} \sum_{i \in PR} 99\, d_i \qquad (B9.10)$$

$$+ \quad \frac{1}{|PX|} \sum_{i \in PX} 77\, (d_i - 47)_+^2 + 88\, d_i \qquad (B9.11)$$

$$+ \quad \frac{1}{|BPL|} \sum_{i \in BPL} 50\, (d_i - 52)_+^2 \qquad (B9.12)$$

$$+ \quad \frac{1}{|BPR|} \sum_{i \in BPR} 110\, (d_i - 60)_+^2 + 66\, d_i \qquad (B9.13)$$

$$+ \quad \frac{1}{|MD|} \sum_{i \in MD} 77\, (d_i - 60)_+^2 + 77\, d_i \qquad (B9.14)$$

$$+ \quad \frac{1}{|SML|} \sum_{i \in SML} 50\, d_i \qquad (B9.15)$$

$$+ \quad \frac{1}{|SMR|} \sum_{i \in SMR} 110\, d_i \qquad (B9.16)$$

$$+ \quad \frac{1}{|NT|} \sum_{i \in NT} 120\, (d_i - d_i^{max})_+^2 \qquad (B9.17)$$

where $SMR$ denotes the set of voxels belonging to the right submandibular gland. For the NTO in Equation (B9.17), the following parameters have been used: $d_0 = 50$ Gy, $d_\infty = 10$ Gy, $x_0 = 0.5$ cm and $\kappa = 0.6$ cm$^{-1}$. In addition to the planning objectives described in Equations (B9.1)-(B9.17), the planning objective in Equation (6) has been used to optimize the fraction-variant treatments for $PTV1$, $PTV2$ and $PTV3$, using similar weights as for the objectives in Equations (B9.1), (B9.2) and (B9.3), respectively.

A non-uniform dose grid size is used, where the size of the smaller voxels is 3.9 x 3.9 x 2.0 mm³. The bixel size is set to 5 x 5 mm² and the photon energy is 6 MV.

### B.10 Patient 10

For patient 10, the objective function in Equation (1) reads as follows:

$$f(\mathbf{d}) \quad = \quad \frac{1}{|PTV1|} \sum_{i \in PTV1} 250\, (52.5 - d_i)_+^2 + 150\, (d_i - 56.3)_+^2 \qquad (B10.1)$$

$$+ \quad \frac{1}{|PTV2|} \sum_{i \in PTV2} 150\, (57 - d_i)_+^2 + 130\, (d_i - 62.8)_+^2 \qquad (B10.2)$$

$$+ \quad \frac{1}{|PTV3|} \sum_{i \in PTV3} 200\, (68.3 - d_i)_+^2 + 190\, (d_i - 72)_+^2 \qquad (B10.3)$$

$$+ \quad \frac{1}{|DT|} \sum_{i \in DT} 88 \, (d_i - 39)_+^2 + 88 \, d_i \qquad (B10.4)$$

$$+ \quad \frac{1}{|BS|} \sum_{i \in B} 99 \, (d_i - 44)_+^2 + 77 \, d_i \qquad (B10.5)$$

$$+ \quad \frac{1}{|SC|} \sum_{i \in SC} 145 \, (d_i - 36)_+^2 \qquad (B10.6)$$

$$+ \quad \frac{1}{|OC|} \sum_{i \in OC} 99 \, (d_i - 51)_+^2 + 88 \, d_i \qquad (B10.7)$$

$$+ \quad \frac{1}{|PL|} \sum_{i \in PL} 135 \, d_i \qquad (B10.8)$$

$$+ \quad \frac{1}{|PR|} \sum_{i \in PR} 120 \, d_i \qquad (B10.9)$$

$$+ \quad \frac{1}{|NT|} \sum_{i \in NT} 100 \, (d_i - d_i^{max})_+^2 \qquad (B10.10)$$

For the NTO in Equation (B10.10), the following parameters have been used: $d_0 = 50$ Gy, $d_\infty = 10$ Gy, $x_0 = 0.5$ cm and $\kappa = 0.6$ cm$^{-1}$. In addition to the planning objectives described in Equations (B10.1)-(B10.10), the planning objective in Equation (6) has been used to optimize the fraction-variant treatments for $PTV1$, $PTV2$ and $PTV3$, using similar weights as for the objectives in Equations (B10.1), (B10.2) and (B10.3), respectively.

A non-uniform dose grid size is used, where the size of the smaller voxels is 3.9 x 3.9 x 2.0 mm³. The bixel size is set to 5 x 5 mm² and the photon energy is 6 MV.

## B.11  Dose calculation algorithm

Calculation of the dose-influence matrix elements $D_{ij}$ is performed with the open-source radiotherapy planning research platform CERR [2] using a quadrant infinite beam (QIB) algorithm [3].

## References Appendix B

# Appendix C  Results

In this section, the dose distributions obtained for patients 2-10 using both fraction-variant and fraction-invariant treatments are reported along with a comparison of the corresponding dose-volume histograms. The dosimetric results achieved for the head-and-neck cancer patients are also summarized.

## C.1  Patient 2

Figure C1 shows the different dose distributions obtained for patient 2 using a fraction-variant treatment that delivers 4 distinct 2-arc VMAT plans in 7 fractions each, while Figure C2 illustrates the results for 2-arc and 3-arc fraction-invariant treatments, along with a comparison of the dose-volume histograms for all plans.

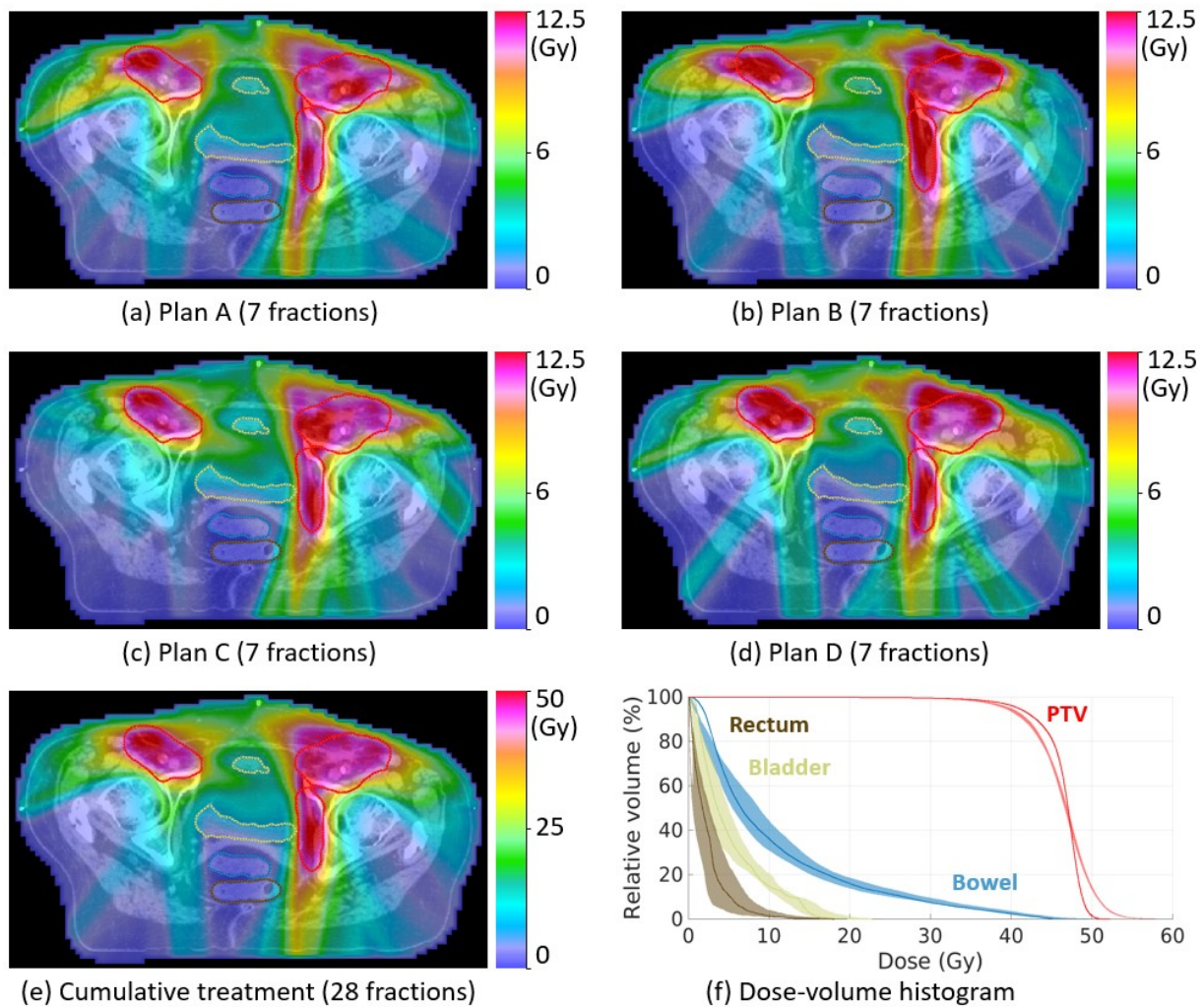

*Figure C1: Fraction-variant treatment obtained for patient 2, which delivers 4 different 2-arc VMAT plans in 7 fractions each (a-d). The dose-volume histogram compares the cumulative treatment resulting from the sum of all individual plans (solid line) with the worst and best case scenarios of the individual plans (assuming that each plan was delivered over 28 fractions; shown by the DVH band).*

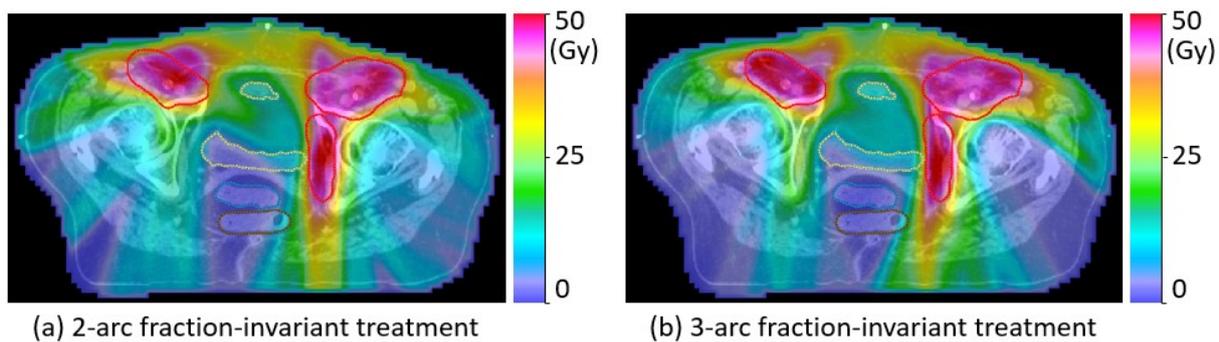
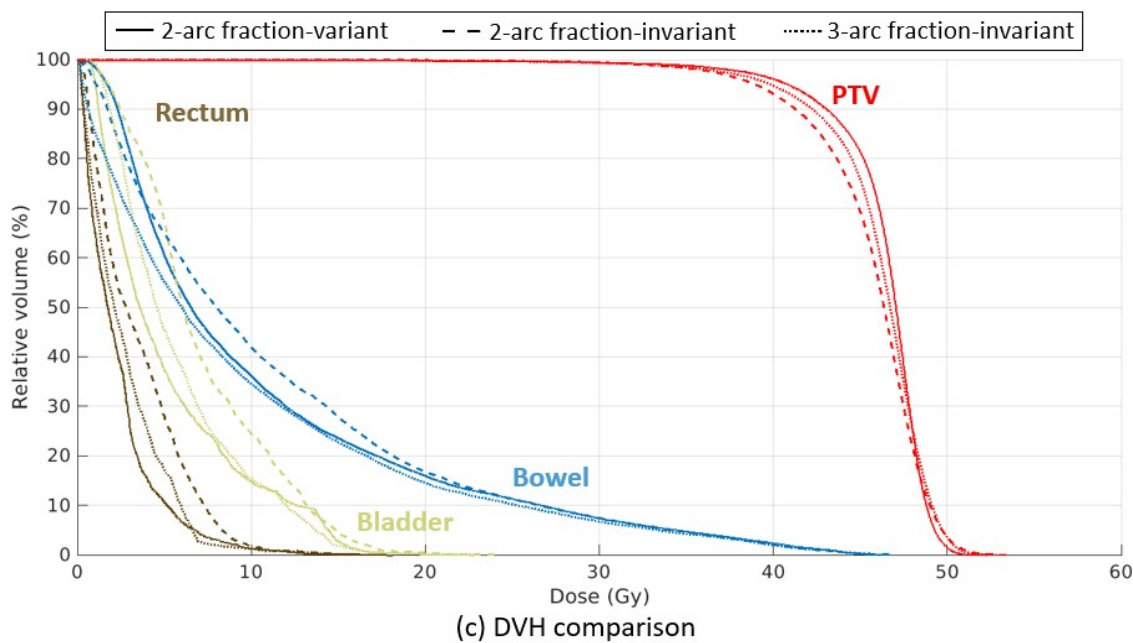

*Figure C2: Dose distributions for the (a) 2-arc and (b) 3-arc fraction-invariant VMAT treatments obtained for patient 2, along with (c) a comparison of the dose-volume histograms between the fraction-variant and fraction-invariant plans.*

## C.2 Patient 3

Figure C3 shows the different dose distributions obtained for patient 3 using a fraction-variant treatment that delivers 5 distinct 2-arc VMAT plans in 5 fractions each, while Figure C4 illustrates the results for 2-arc and 3-arc fraction-invariant treatments, along with a comparison of the dose-volume histograms for all plans.

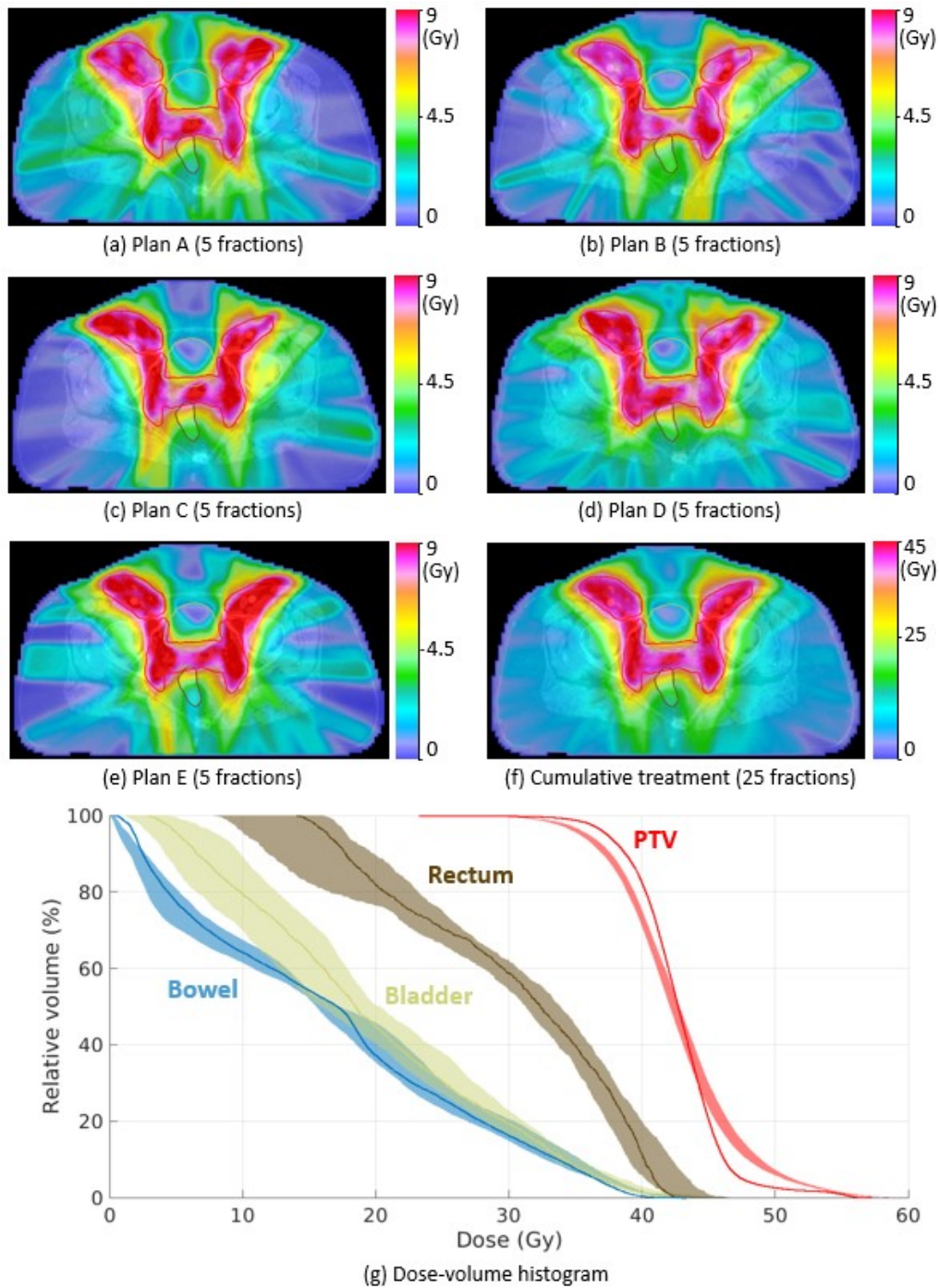

*Figure C3: Fraction-variant treatment obtained for patient 3, which delivers 5 different 2-arc VMAT plans in 5 fractions each (a-e). The dose-volume histogram compares the cumulative treatment resulting from the sum of all individual plans (solid line) with the worst and best case scenarios of the individual plans (assuming that each plan was delivered over 25 fractions; shown by the DVH band).*

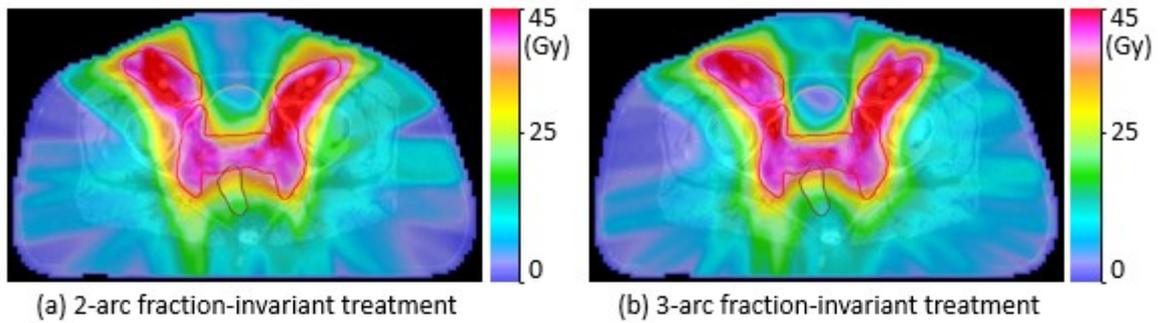
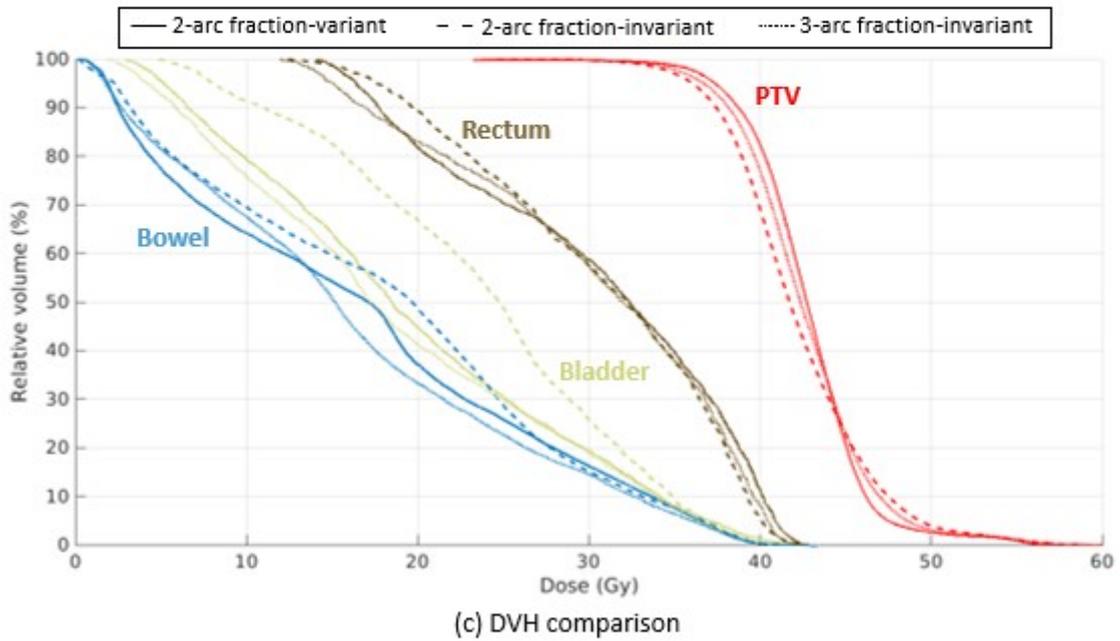

*Figure C4: Dose distributions for the (a) 2-arc and (b) 3-arc fraction-invariant VMAT treatments obtained for patient 3, along with (c) a comparison of the dose-volume histograms between the fraction-variant and fraction-invariant plans.*

## C.3 Patient 4

Figure C5 shows the different dose distributions obtained for patient 4 using a fraction-variant treatment that delivers 5 distinct 2-arc VMAT plans in 5 fractions each, while Figure C6 illustrates the results for 2-arc and 3-arc fraction-invariant treatments, along with a comparison of the dose-volume histograms for all plans.

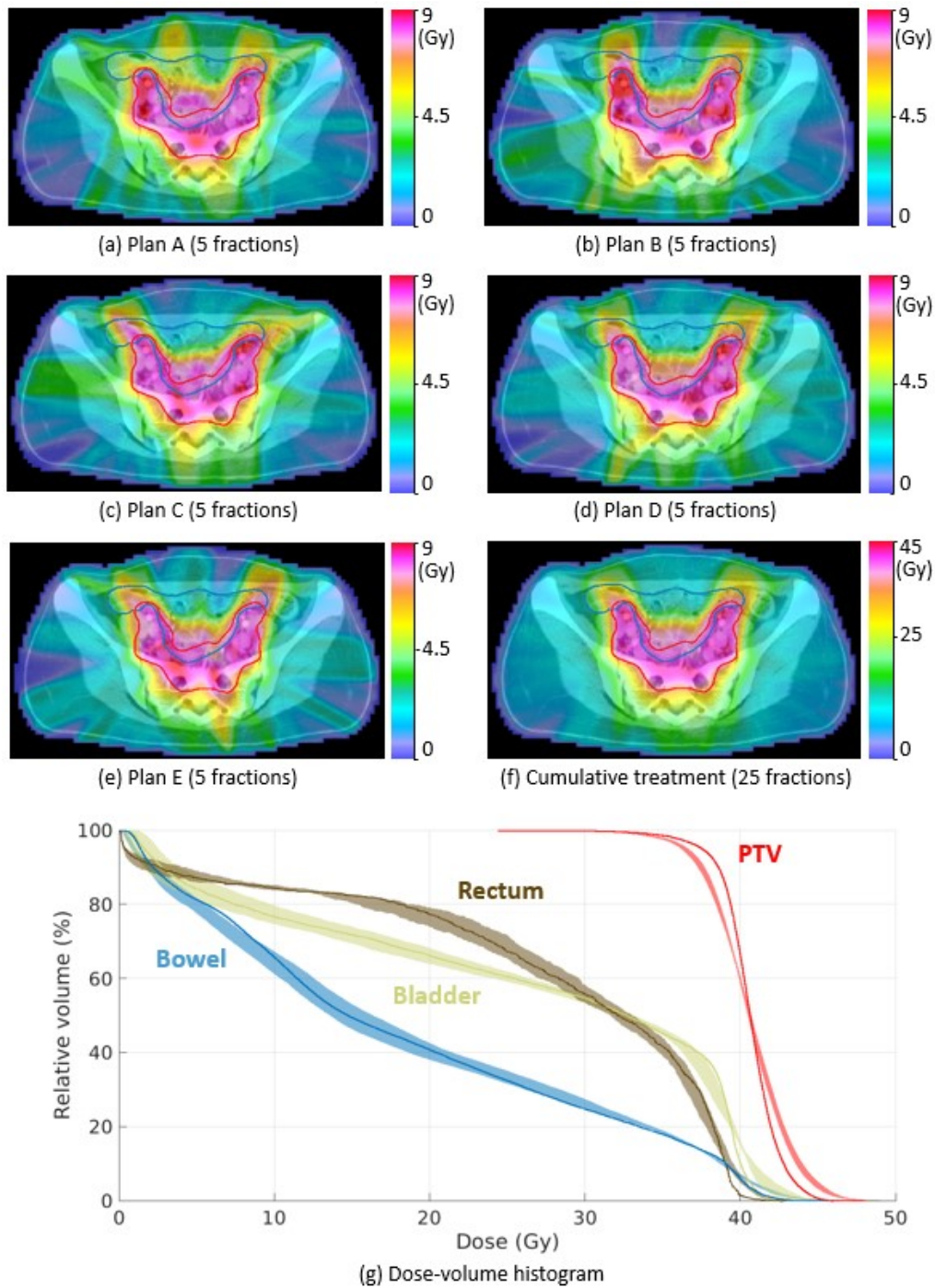

*Figure C5: Fraction-variant treatment obtained for patient 4, which delivers 5 different 2-arc VMAT plans in 5 fractions each (a-e). The dose-volume histogram compares the cumulative treatment resulting from the sum of all individual plans (solid line) with the worst and best case scenarios of the individual plans (assuming that each plan was delivered over 25 fractions; shown by the DVH band).*

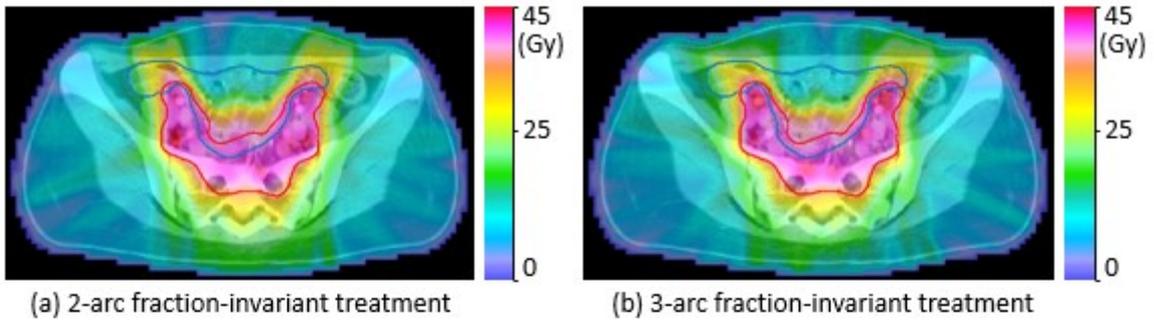
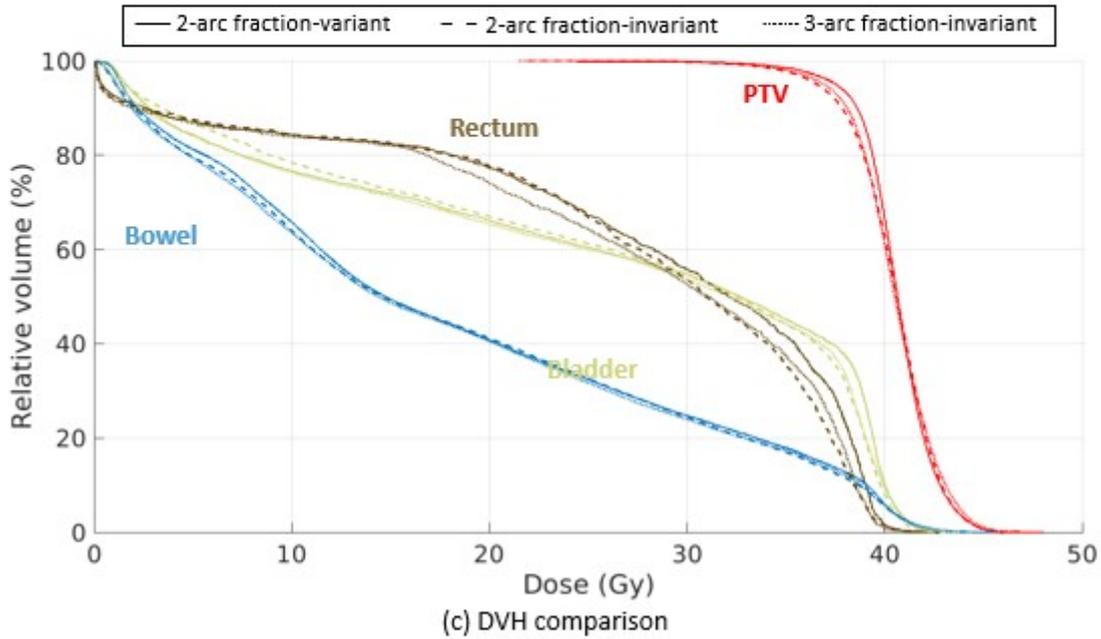

*Figure C6: Dose distributions for the (a) 2-arc and (b) 3-arc fraction-invariant VMAT treatments obtained for patient 4, along with (c) a comparison of the dose-volume histograms between the fraction-variant and fraction-invariant plans.*

## C.4 Patient 5

Figure C7 shows the different dose distributions obtained for patient 5 using a fraction-variant treatment that delivers 7 distinct 2-arc VMAT plans in 2 fractions each, while Figure C8 illustrates the results for 2-arc and 3-arc fraction-invariant treatments, along with a comparison of the dose-volume histograms for all plans.

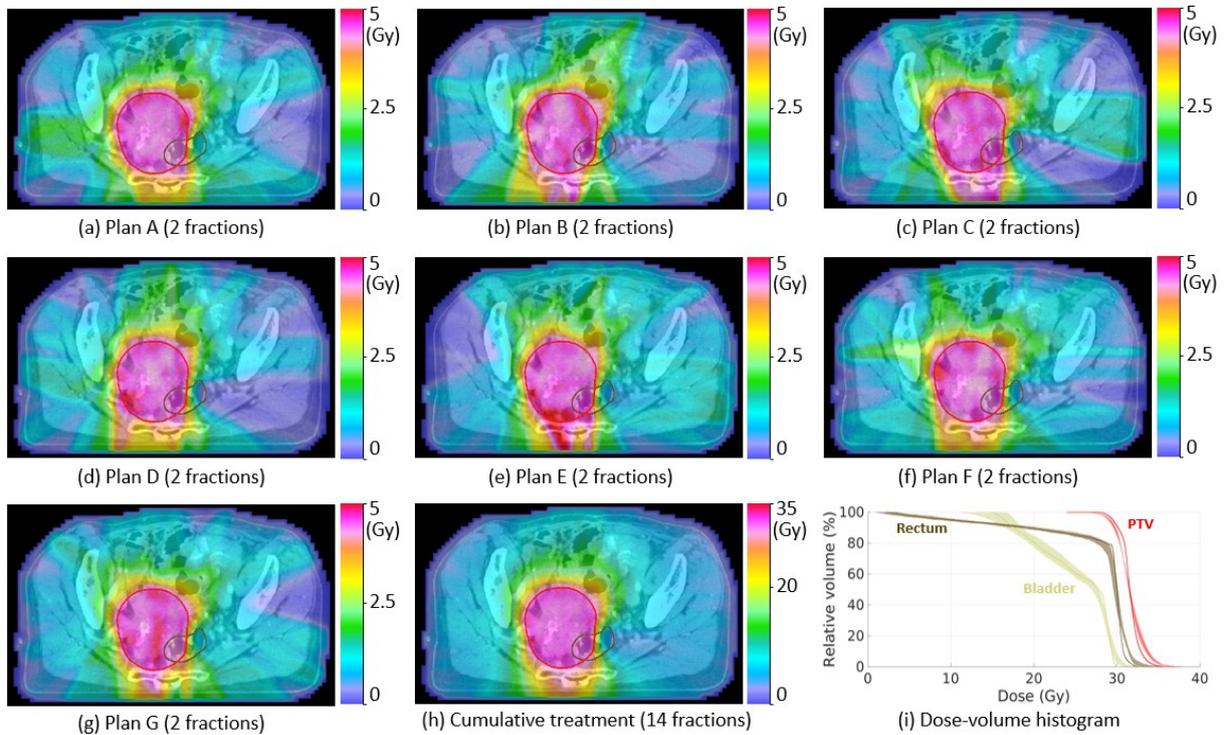

*Figure C7: Fraction-variant treatment obtained for patient 5, which delivers 7 different 2-arc VMAT plans in 2 fractions each (a-g). The dose-volume histogram compares the cumulative treatment resulting from the sum of all individual plans (solid line) with the worst and best case scenarios of the individual plans (assuming that each plan was delivered over 14 fractions; shown by the DVH band).*

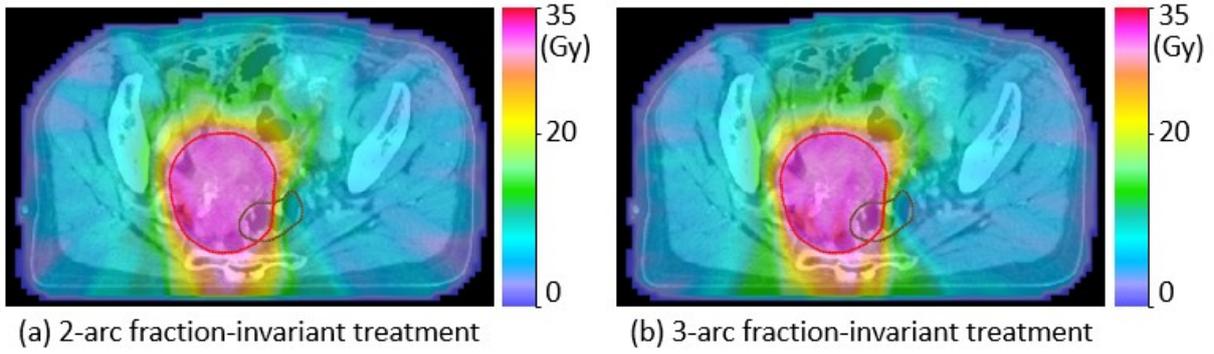
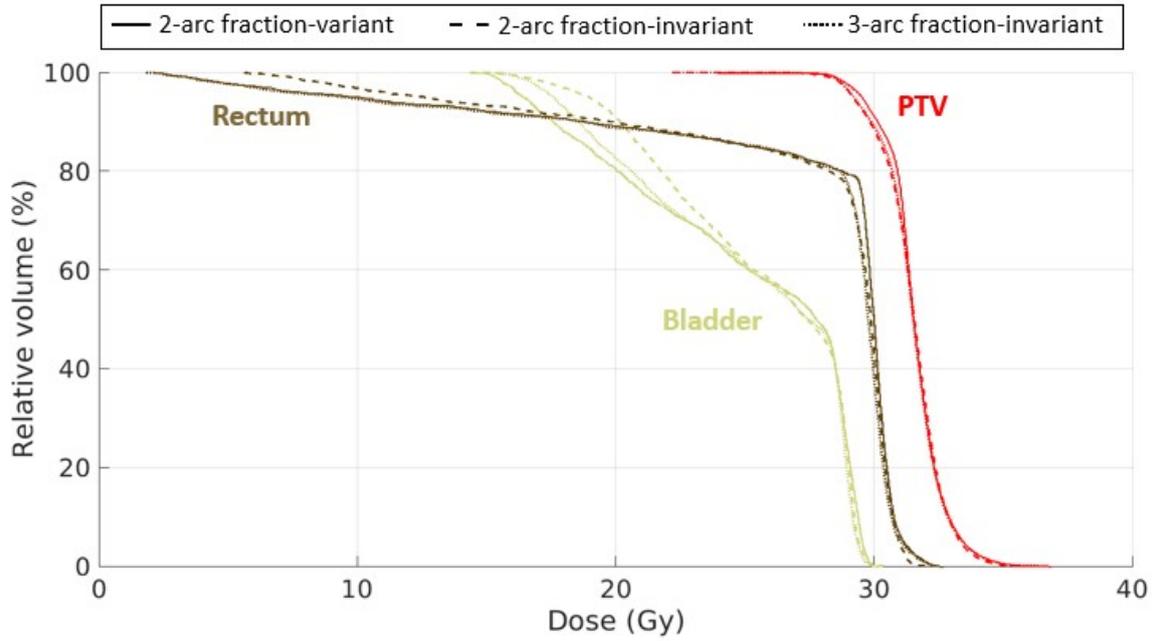

*Figure C8: Dose distributions for the (a) 2-arc and (b) 3-arc fraction-invariant VMAT treatments obtained for patient 5, along with (c) a comparison of the dose-volume histograms between the fraction-variant and fraction-invariant plans.*

## C.5  Patient 6

Figure C9 shows the different dose distributions obtained for patient 6 using a fraction-variant treatment that delivers 5 distinct 2-arc VMAT plans in 7 fractions each, while Figure C10 illustrates the results for 2-arc and 3-arc fraction-invariant treatments, along with a comparison of the dose-volume histograms for all plans.

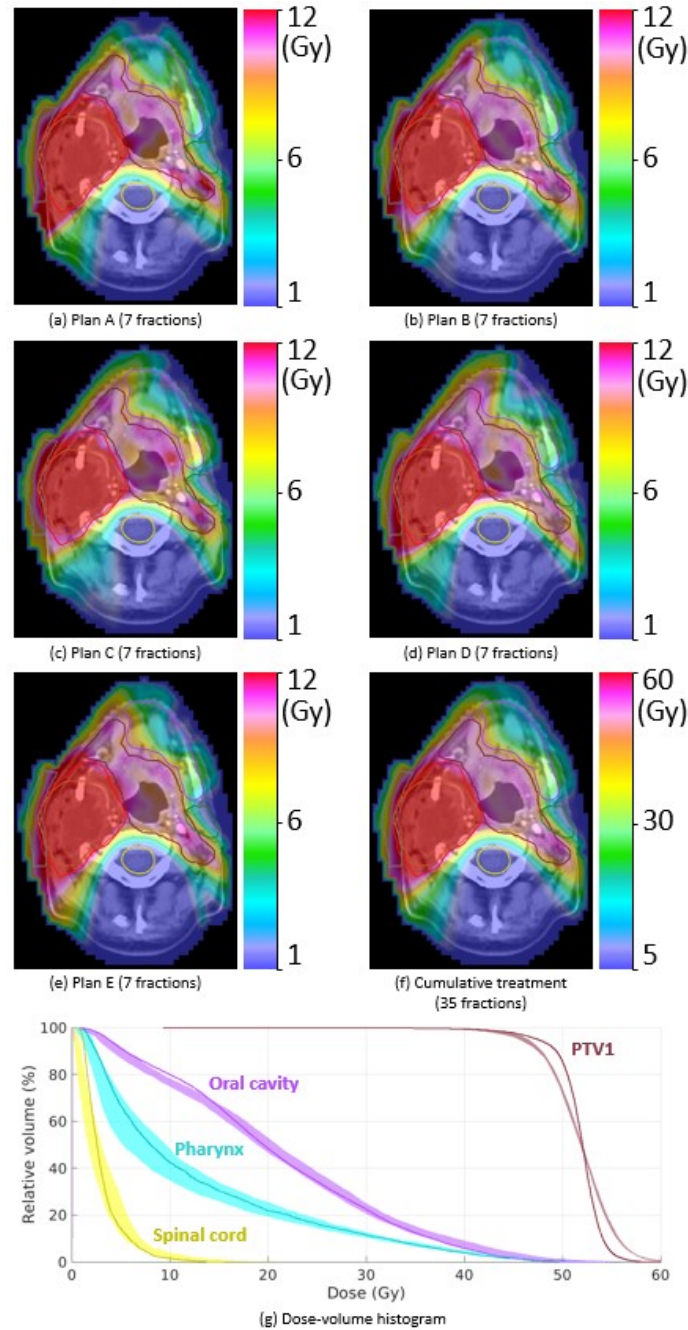

*Figure C9: Fraction-variant treatment obtained for patient 6, which delivers 5 different 2-arc VMAT plans in 7 fractions each (a-e). The dose-volume histogram compares the cumulative treatment resulting from the sum of all individual plans (solid line) with the worst and best case scenarios of the individual plans (assuming that each plan was delivered over 35 fractions; shown by the DVH band). Contours of PTV1 (dark red), PTV2 (red), PTV3 (orange), spinal cord (yellow), oral cavity (purple), parotid glands (green) and a 10 mm flab (grey) are visible.*

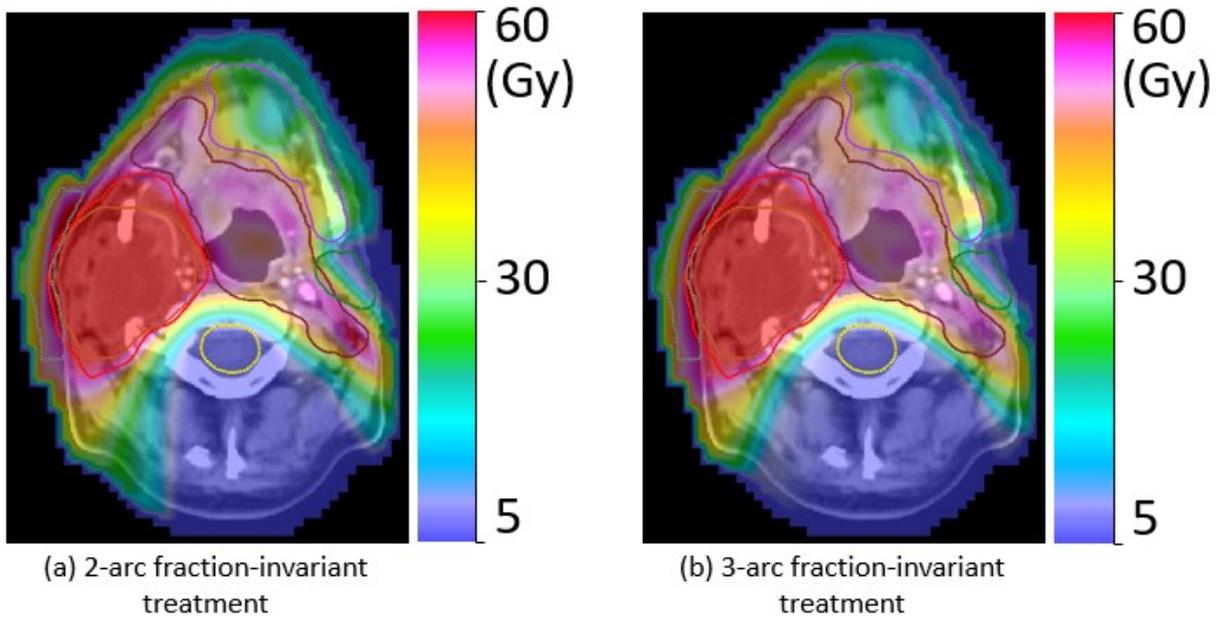
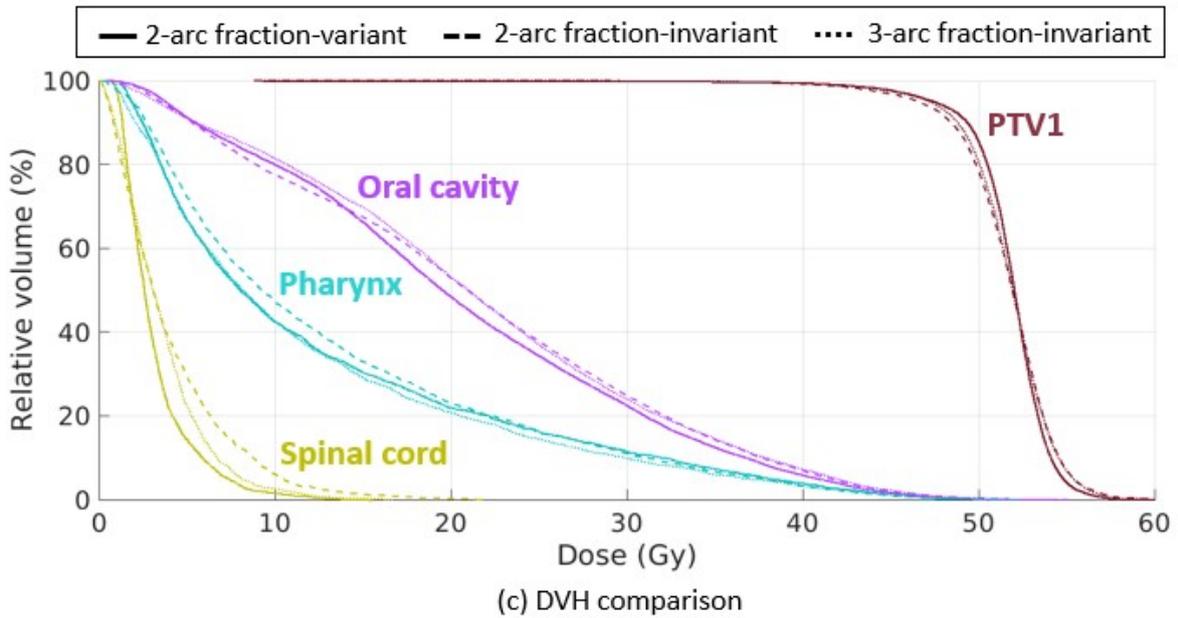

Figure C10: Dose distributions for the (a) 2-arc and (b) 3-arc fraction-invariant VMAT treatments obtained for patient 6, along with (c) a comparison of the dose-volume histograms between the fraction-variant and fraction-invariant plans.

## C.6 Patient 7

Figure C11 shows the different dose distributions obtained for patient 7 using a fraction-variant treatment that delivers 5 distinct 2-arc VMAT plans in 7 fractions each, while Figure C12 illustrates the results for 2-arc and 3-arc fraction-invariant treatments, along with a comparison of the dose-volume histograms for all plans.

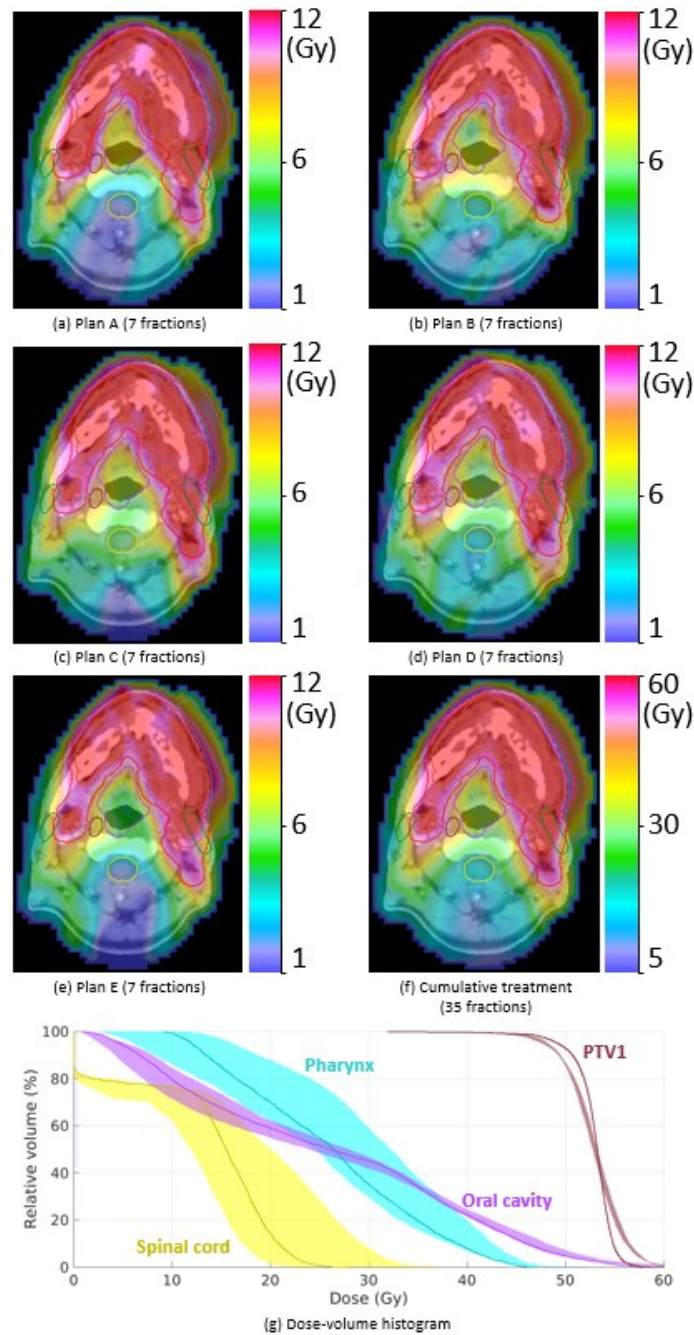

*Figure C11: Fraction-variant treatment obtained for patient 7, which delivers 5 different 2-arc VMAT plans in 7 fractions each (a-e). The dose-volume histogram compares the cumulative treatment resulting from the sum of all individual plans (solid line) with the worst and best case scenarios of the individual plans (assuming that each plan was delivered over 35 fractions; shown by the DVH band). Contours of PTV2 (red), PTV3 (orange), spinal cord (yellow), oral cavity (purple), parotid glands (green) and a 10 mm flab (grey) are visible.*

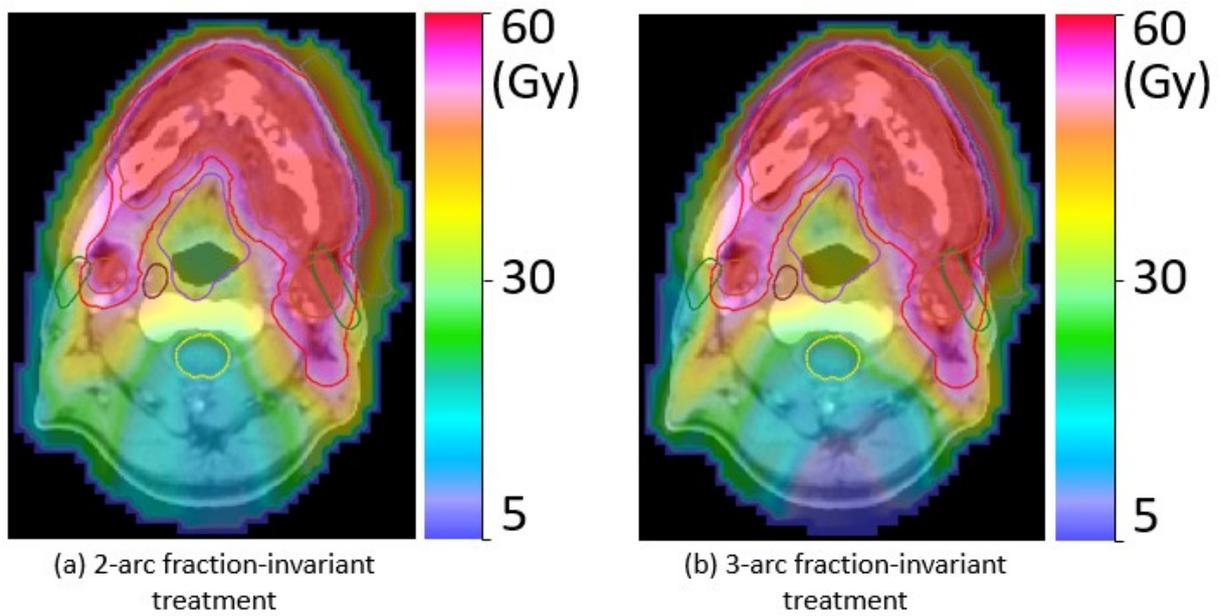
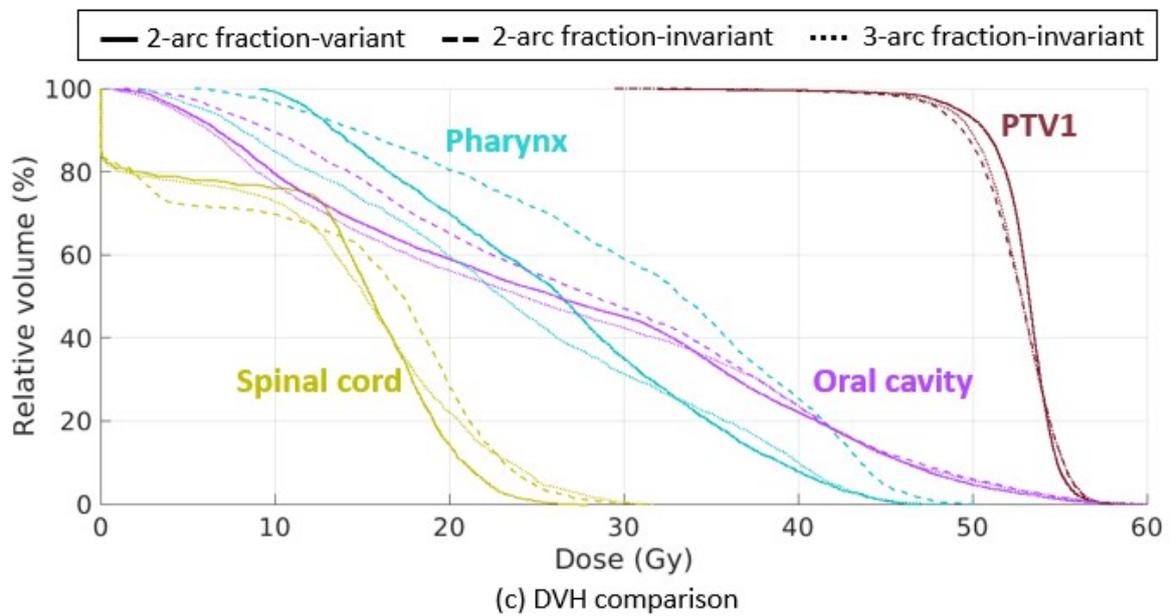

*Figure C12: Dose distributions for the (a) 2-arc and (b) 3-arc fraction-invariant VMAT treatments obtained for patient 7, along with (c) a comparison of the dose-volume histograms between the fraction-variant and fraction-invariant plans.*

## C.7 Patient 8

Figure C13 shows the different dose distributions obtained for patient 8 using a fraction-variant treatment that delivers 5 distinct 2-arc VMAT plans in 7 fractions each, while Figure C14 illustrates the results for 2-arc and 3-arc fraction-invariant treatments, along with a comparison of the dose-volume histograms for all plans.

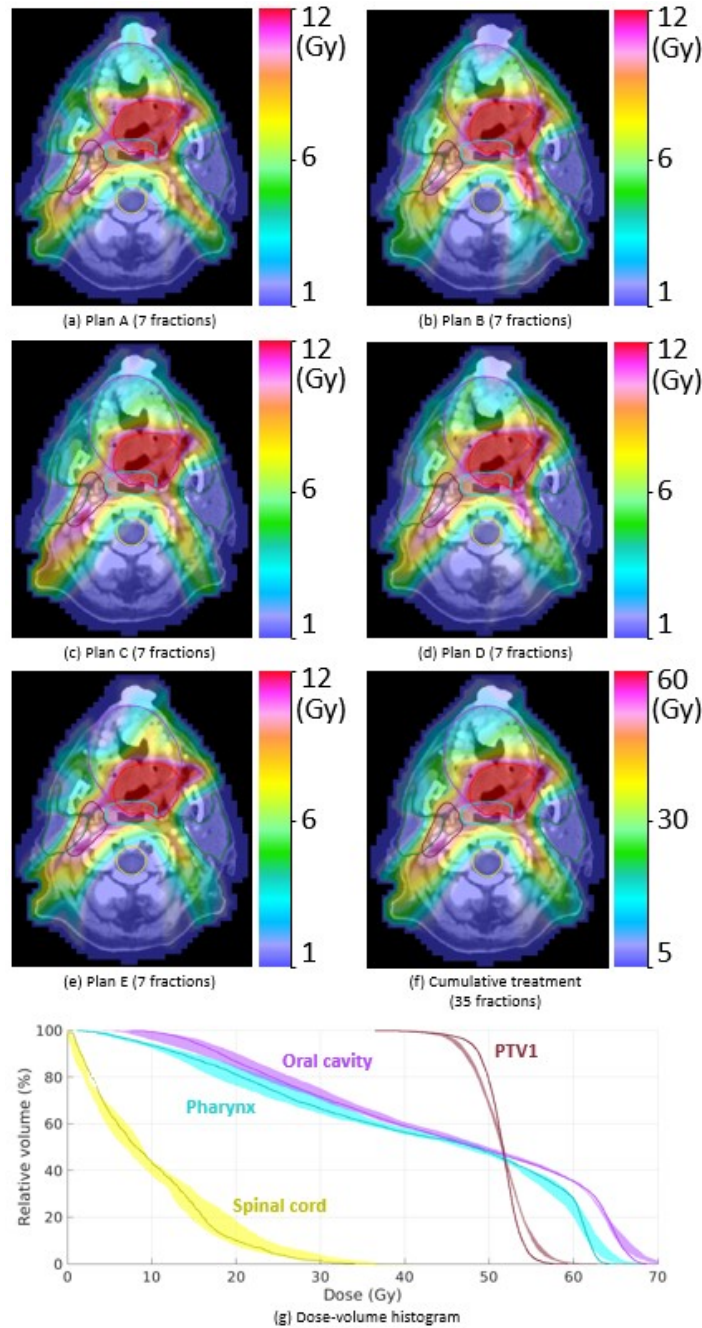

*Figure C13: Fraction-variant treatment obtained for patient 8, which delivers 5 different 2-arc VMAT plans in 7 fractions each (a-e). The dose-volume histogram compares the cumulative treatment resulting from the sum of all individual plans (solid line) with the worst and best case scenarios of the individual plans (assuming that each plan was delivered over 35 fractions; shown by the DVH band). Contours of PTV1 (dark red), spinal cord (yellow), oral cavity (purple), parotid glands (green) and pharynx (light blue) are visible.*

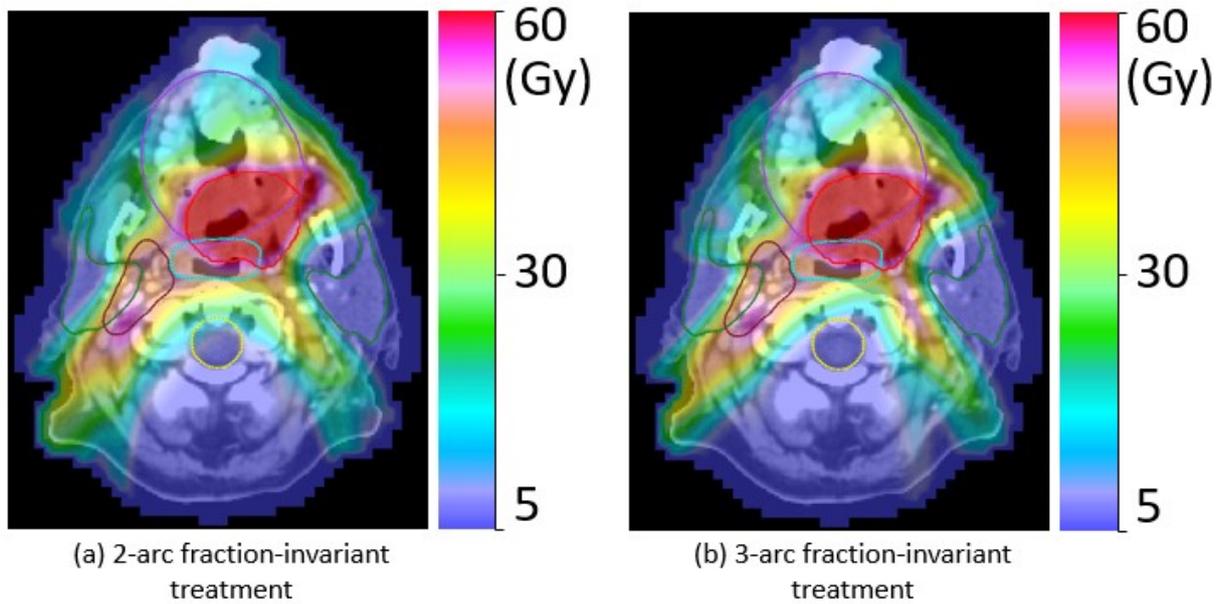

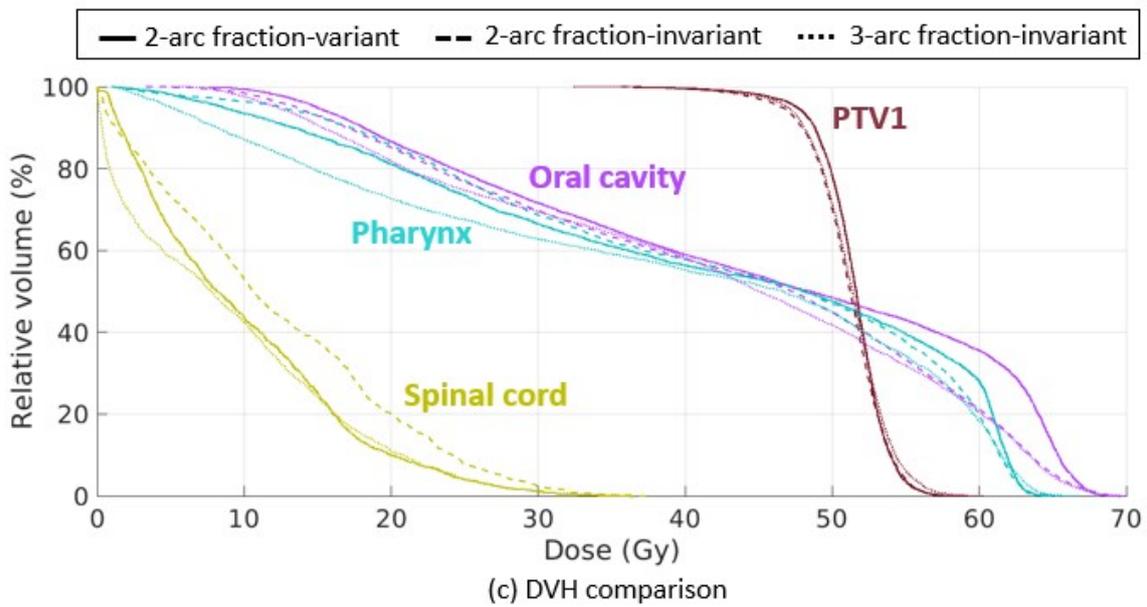

*Figure C14: Dose distributions for the (a) 2-arc and (b) 3-arc fraction-invariant VMAT treatments obtained for patient 8, along with (c) a comparison of the dose-volume histograms between the fraction-variant and fraction-invariant plans.*

## C.8  Patient 9

Figure C15 shows the different dose distributions obtained for patient 9 using a fraction-variant treatment that delivers 5 distinct 2-arc VMAT plans in 7 fractions each, while Figure C16 illustrates the results for 2-arc and 3-arc fraction-invariant treatments, along with a comparison of the dose-volume histograms for all plans.

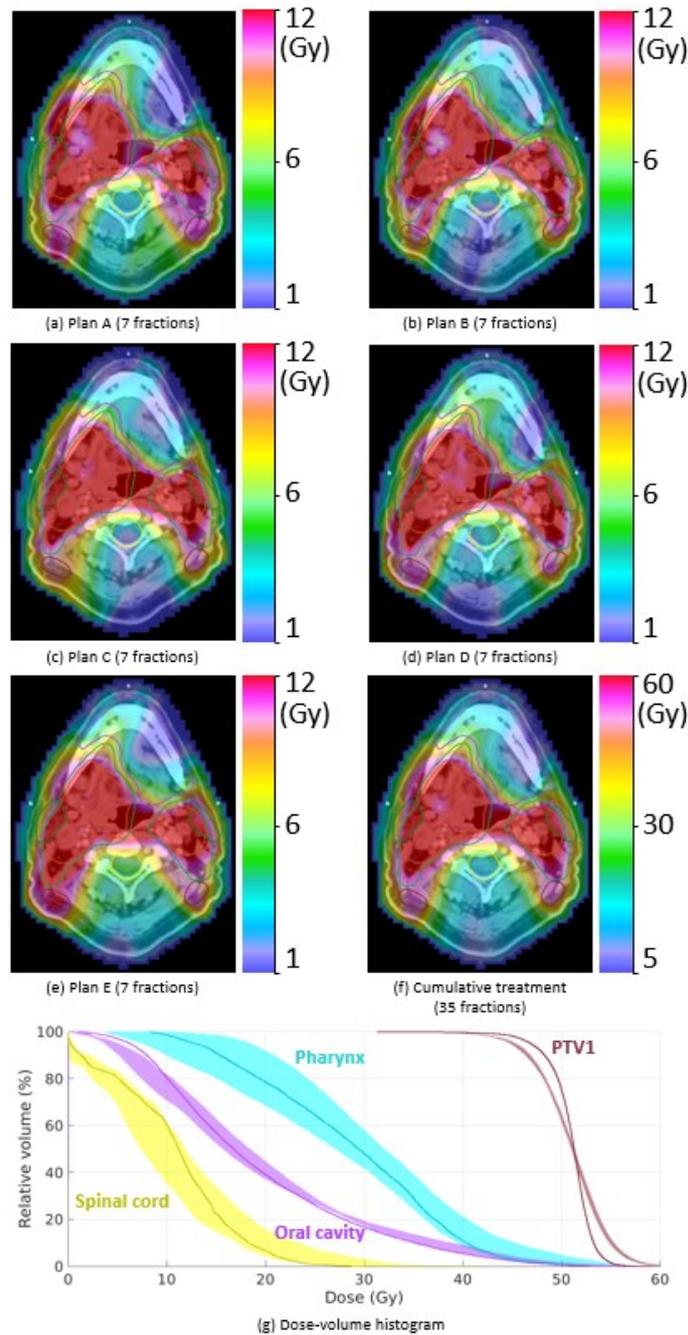

*Figure C15: Fraction-variant treatment obtained for patient 9, which delivers 5 different 2-arc VMAT plans in 7 fractions each (a-e). The dose-volume histogram compares the cumulative treatment resulting from the sum of all individual plans (solid line) with the worst and best case scenarios of the individual plans (assuming that each plan was delivered over 35 fractions; shown by the DVH band). Contours of PTV1 (dark red), PTV2 (red), PTV3 (orange), spinal cord (yellow), oral cavity (purple) and parotid glands (green) are visible.*

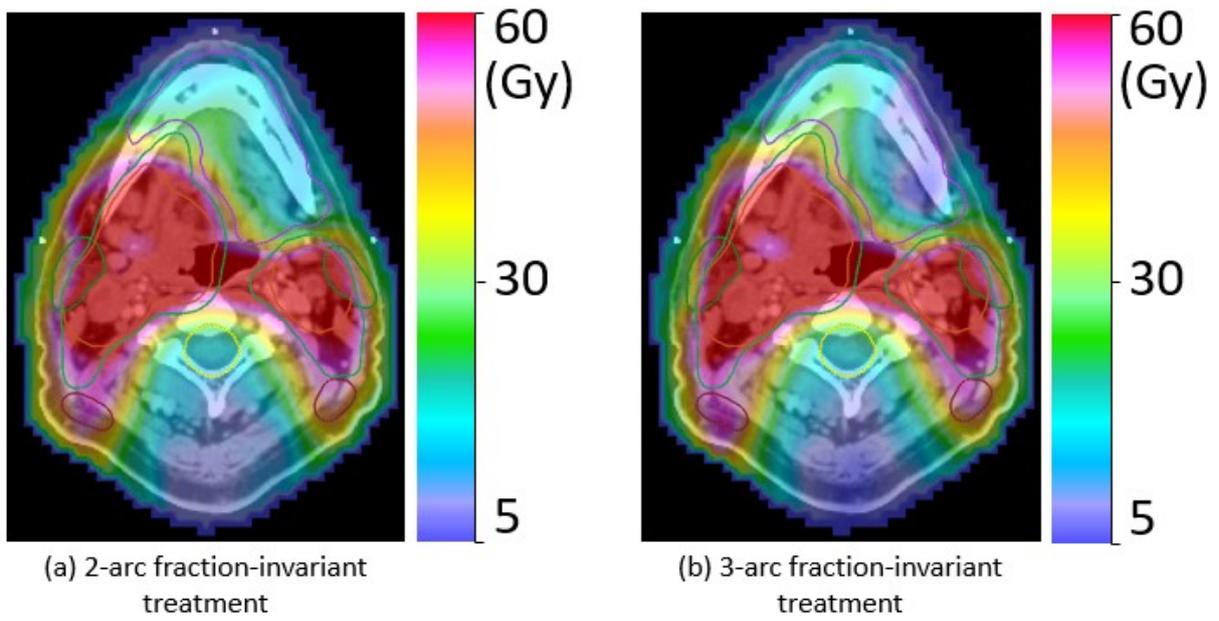
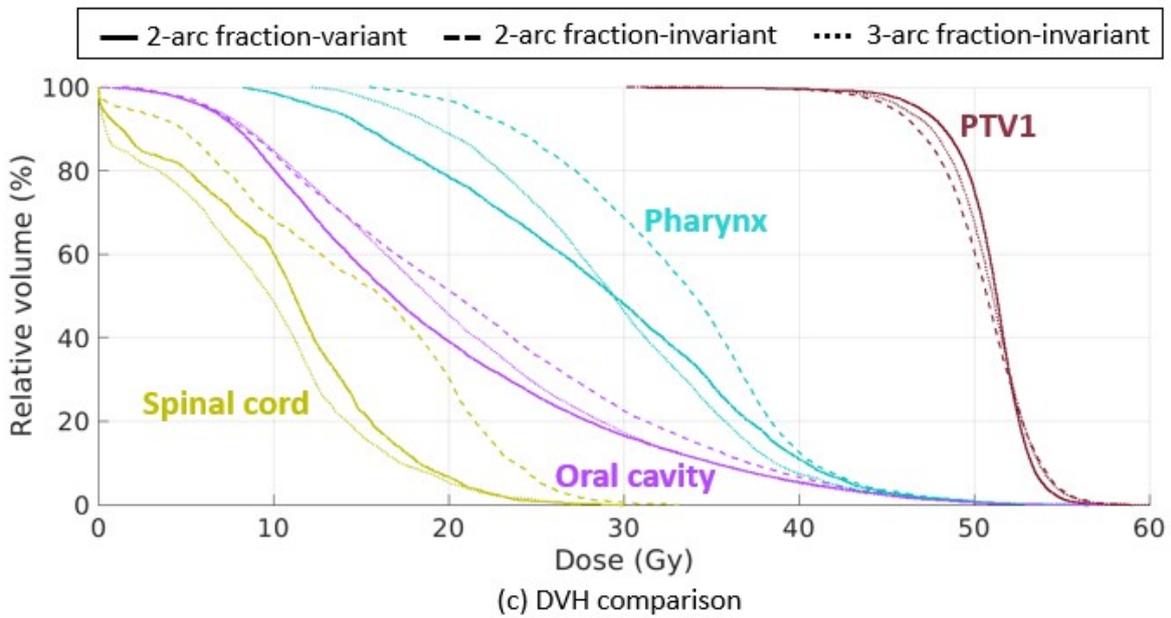

*Figure C16: Dose distributions for the (a) 2-arc and (b) 3-arc fraction-invariant VMAT treatments obtained for patient 9, along with (c) a comparison of the dose-volume histograms between the fraction-variant and fraction-invariant plans.*

## C.9  Patient 10

Figure C17 shows the different dose distributions obtained for patient 10 using a fraction-variant treatment that delivers 5 distinct 2-arc VMAT plans in 7 fractions each, while Figure C18 illustrates the results for 2-arc and 3-arc fraction-invariant treatments, along with a comparison of the dose-volume histograms for all plans.

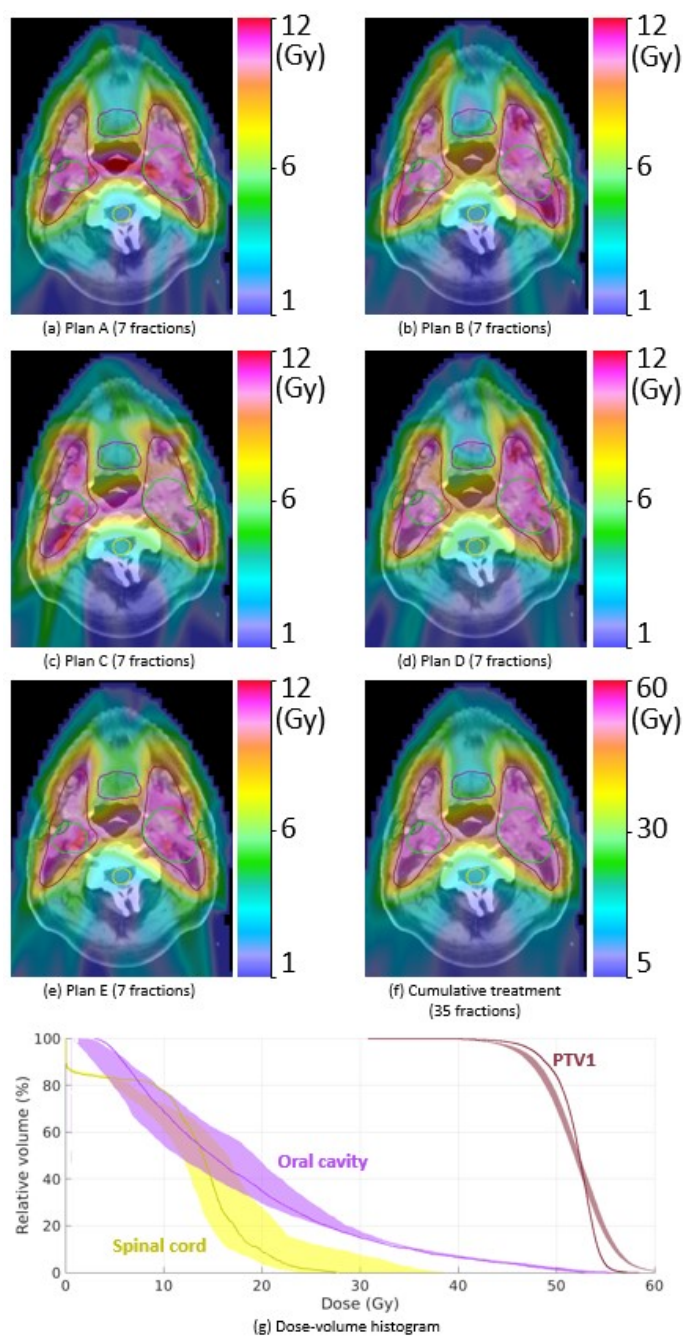

*Figure C17: Fraction-variant treatment obtained for patient 10, which delivers 5 different 2-arc VMAT plans in 7 fractions each (a-e). The dose-volume histogram compares the cumulative treatment resulting from the sum of all individual plans (solid line) with the worst and best case scenarios of the individual plans (assuming that each plan was delivered over 35 fractions; shown by the DVH band).*

*Contours of PTV1 (dark red), PTV2 (light green), spinal cord (yellow), oral cavity (purple), parotid glands (green) and a 10 mm flab (grey) are visible.*

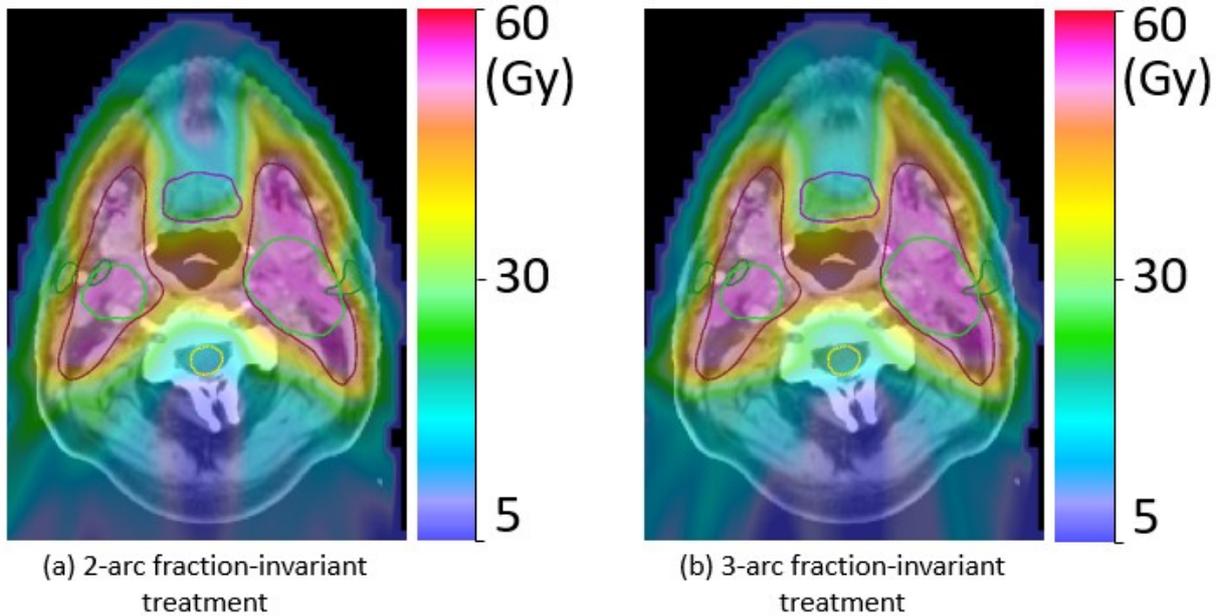

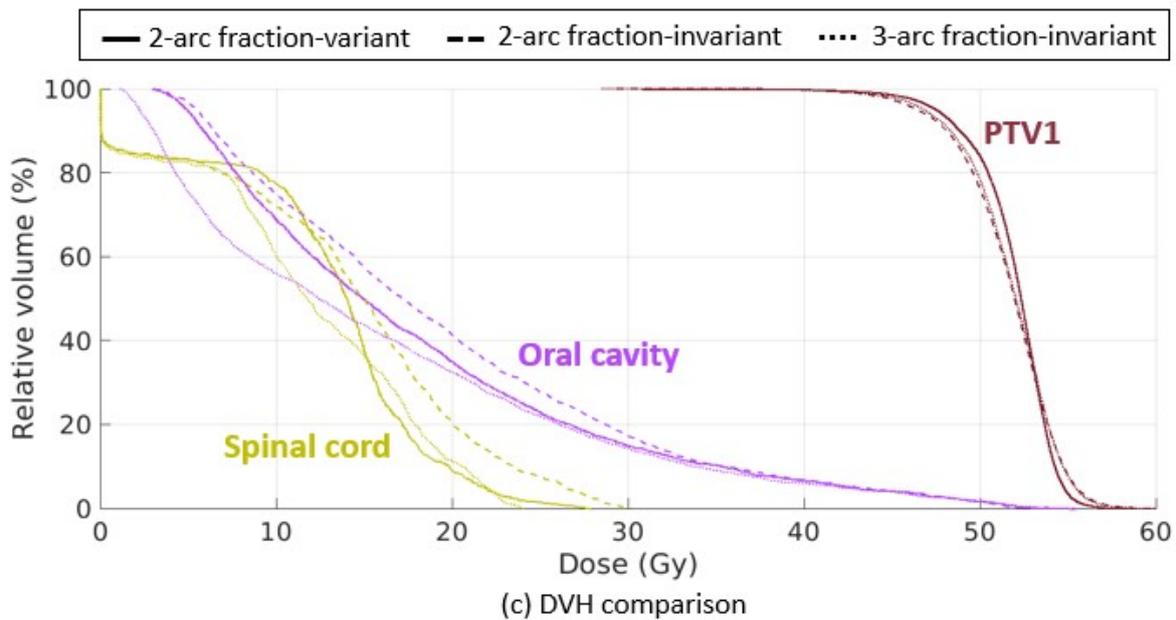

*Figure C17: Dose distributions for the (a) 2-arc and (b) 3-arc fraction-invariant VMAT treatments obtained for patient 10, along with (c) a comparison of the dose-volume histograms between the fraction-variant and fraction-invariant plans.*